\documentclass[prl,amsmath,twocolumn,amssymb,aps,superscriptaddress,floatfix,preprintnumbers,10pt,notitlepage,openany]{revtex4-1}
\usepackage{times}
\usepackage{graphicx}
\usepackage{amsmath, braket, amsfonts}
\usepackage{amssymb}
\usepackage{natbib}
\usepackage{sidecap}
\usepackage{bm, color, ulem}
\usepackage{xcolor}
\usepackage{lipsum}
\usepackage{changes}
\usepackage{titlesec}

\usepackage{multirow}
\usepackage{makecell}

\titlespacing{\section}
{0pt}{2.5ex plus .2ex minus .2ex}{2.5ex plus .2ex minus .2ex}

\titlespacing{\subsection}
{0pt}{1.5ex plus .2ex minus .2ex}{1.5ex plus .2ex minus .2ex}

\begin{document}

\title{Interface Piezoelectric Loss in Superconducting Qubits}

\author{Haoxin Zhou}
\thanks{H.Z. and K.Y. contributed equally.}
\affiliation{
Department of Electrical Engineering and Computer Sciences, University of California,  Berkeley, Berkeley, California 94720, USA
}
\affiliation{
 Materials Sciences Division, Lawrence Berkeley National Laboratory, Berkeley, California 94720, USA
}
\affiliation{
Department of Physics, University of California, Berkeley, Berkeley, California 94720, USA
}

\author{Kangdi Yu}
\thanks{H.Z. and K.Y. contributed equally.}
\affiliation{
Department of Electrical Engineering and Computer Sciences, University of California,  Berkeley, Berkeley, California 94720, USA
}
\affiliation{
 Materials Sciences Division, Lawrence Berkeley National Laboratory, Berkeley, California 94720, USA
}
\author{Yashwanth Balaji}

\affiliation{
Molecular Foundry, Lawrence Berkeley National Laboratory, Berkeley, California 94720, USA
}
\author{Sanjit Shirol}
\affiliation{
Department of Physics, University of California,  Berkeley, Berkeley, California 94720, USA
}

\author{Leo Sementilli}
\affiliation{
Department of Electrical Engineering and Computer Sciences, University of California,  Berkeley, Berkeley, California 94720, USA
}
\affiliation{
 Materials Sciences Division, Lawrence Berkeley National Laboratory, Berkeley, California 94720, USA
}
\affiliation{
Department of Physics, University of California, Berkeley, Berkeley, California 94720, USA
}

\author{Zi-Huai Zhang}
\affiliation{
Department of Electrical Engineering and Computer Sciences, University of California,  Berkeley, Berkeley, California 94720, USA
}
\affiliation{
 Materials Sciences Division, Lawrence Berkeley National Laboratory, Berkeley, California 94720, USA
}
\affiliation{
Department of Physics, University of California, Berkeley, Berkeley, California 94720, USA
}

\author{Adam Schwartzberg}

\affiliation{
Molecular Foundry, Lawrence Berkeley National Laboratory, Berkeley, California 94720, USA
}

\author{Alp Sipahigil}
\email{Corresponding author: alp@berkeley.edu}

\affiliation{
Department of Electrical Engineering and Computer Sciences, University of California,  Berkeley, Berkeley, California 94720, USA
}
\affiliation{
 Materials Sciences Division, Lawrence Berkeley National Laboratory, Berkeley, California 94720, USA
}
\affiliation{
Department of Physics, University of California, Berkeley, Berkeley, California 94720, USA
}

\begin{abstract}
Dissipation remains a central obstacle to improving superconducting quantum circuits, yet the microscopic origins of loss in widely used materials platforms are not fully understood.
Here, we report the observation of interface piezoelectricity-induced dissipation in superconducting qubits fabricated on high-resistivity silicon.
Our devices use a transmon qubit with a shunt capacitor that simultaneously serves as an interdigital transducer embedded in a surface acoustic wave resonator.
By tuning the qubit transition into resonance with discrete mechanical modes, we observe up to a factor-of-two reduction in qubit lifetime, consistent with energy exchange between the qubit and mechanical modes mediated by piezoelectric coupling at the aluminum-silicon interface.
Our findings provide direct evidence for interface piezoelectricity as a distinct loss channel in superconducting qubits. Combined with multiphysics simulations, these findings suggest that interface piezoelectric loss can dominate over loss from two-level systems at sufficiently high frequencies.
\end{abstract}

\maketitle

Identifying the mechanisms that limit macroscopic quantum coherence in superconducting qubits has become central to scalable, fault-tolerant quantum information processing and to the broader study of dissipation in solids~\cite{oliver_materials_2013, de_leon_materials_2021}.
Qubit losses are commonly attributed to material disorder, often modeled as two-level systems (TLSs) in dielectrics that couple to microwave fields and absorb energy~\cite{simmonds_decoherence_2004, muller_towards_2019}.
Continued advances in materials and fabrication have substantially reduced such disorder-induced losses~\cite{bland_2d_2025}.
Yet, additional dissipation channels may arise from intrinsic or structural properties of the systems themselves.

Electromechanical coupling is one possible intrinsic dissipation pathway for superconducting qubits.
In strongly piezoelectric substrates, microwave electric fields can couple efficiently to acoustic phonons, allowing superconducting qubits to radiate energy into propagating acoustic modes~\cite{gustafsson_propagating_2014}.
Such phonon-mediated relaxation has been observed in superconducting qubits fabricated on or in proximity to piezoelectric materials \cite{satzinger_quantum_2018, jain_acoustic_2023}.
Piezoelectric loss can be nominally avoided by using non-piezoelectric substrates such as silicon or sapphire, where the centrosymmetry of their crystal structures suppresses electromechanical dissipation~\cite{ashcroft_solid_1976}. Recent classical electromechanical measurements have shown that superconductor-substrate interfaces exhibit an effective piezoelectric response even when the bulk material is centrosymmetric~\cite{zhou_observation_2025}. Whether this coupling limits the coherence of superconducting qubits, and how it scales with device geometry and frequency, has remained an open question: phonon emission from a qubit on a centrosymmetric substrate has not been directly measured.

Here, we present direct evidence that interface piezoelectricity causes superconducting qubits on silicon to radiate energy into propagating phonons. This is achieved by coupling a frequency-tunable transmon qubit to a surface acoustic wave (SAW) resonator.
When the qubit is tuned into resonance with the acoustic modes, its lifetime is significantly reduced, consistent with energy loss through phonon radiation~\cite{jain_acoustic_2023}.
These findings establish interface piezoelectricity as a previously overlooked source of dissipation~\cite{siddiqi_engineering_2021} that can intrinsically limit qubit lifetime, even when the qubits are fabricated on nominally non-piezoelectric materials. 
Moreover, our findings reveal how symmetry breaking at material interfaces enables unexpected couplings between quantum devices and propagating phonons, opening new opportunities to engineer coherent qubit-phonon interactions.

\begin{figure}[!t]
\centering
\includegraphics[width=1.0\columnwidth]{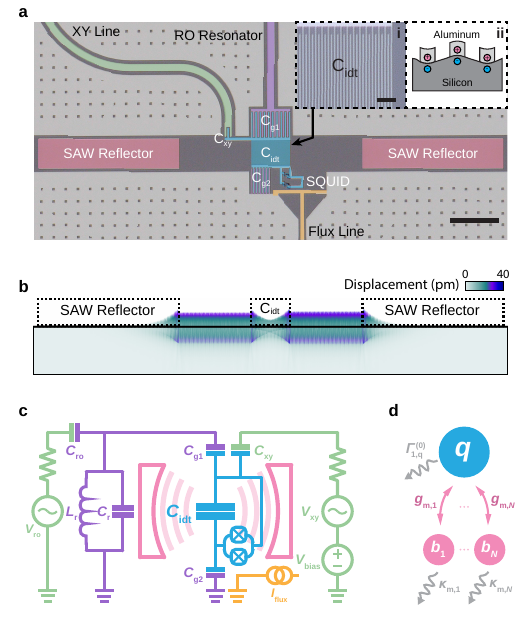}
\vspace{-6mm}
\caption{\textbf{A transmon qubit inside a SAW resonator on silicon. }
\textbf{a,} False-colored optical micrograph of a device studied (Sample C). A floating transmon qubit (blue), formed by a SQUID shunted by an IDT capacitor $C_{\text{idt}}$, couples capacitively to a quarter-wavelength readout resonator (purple) via $C_{\text{g1}}$ and $C_{\text{g2}}$. The qubit is driven through the XY line (green) and flux-tuned via the flux line (yellow). A SAW resonator, defined by two reflectors, spatially overlaps with the qubit electric field, enabling piezoelectric electromechanical coupling. 
Left inset: Enlarged view of the IDT fingers. Scale bars: $50~\mu\text{m}$ (main), $5~\mu\text{m}$ (inset).
Right inset: Schematic illustration of the interface-mediated piezoelectric response.
\textbf{b,} Side view of the simulated mechanical displacement field for a representative SAW resonator mode, showing evanescent decay into the reflector regions.
\textbf{c,} Lumped-element circuit model. The mirrors (pink) represent the SAW resonator. The XY line is connected to a bias tee combining a DC bias $V_{\text{bias}}$ and an AC drive $V_{\text{xy}}$. The readout tone $V_{\text{ro}}$ couples to the resonator, and the flux bias $I_{\text{flux}}$ tunes the SQUID.
\textbf{d,} Simplified quantum-optical model of the piezoelectric interaction. The qubit couples to each SAW resonator mode with strength $g_{\text{m},k}$. Here, $\Gamma_{1,\text{q}}^{(0)}$ is the intrinsic qubit loss of non-piezoelectric origin, and $\kappa_{\text{m},k}$ are the SAW resonator mode decay rates.
}
\label{fig:fig1}
\end{figure}
    
\section{Interface piezoelectric qubit-phonon coupling}

To probe qubit-phonon coupling mediated by interface piezoelectricity, we integrate a superconducting transmon qubit directly into a SAW resonator~\cite{manenti_the_2017, Moores_IDT} (Fig.~\ref{fig:fig1}a).
The device is fabricated by patterning a 50-nm-thick aluminum film on an intrinsic silicon substrate.
The qubit consists of a superconducting interference device (SQUID) shunted by an interdigital capacitor $C_{\text{idt}}$, which is positioned between two Bragg reflectors that define the SAW resonator.
In this geometry, $C_{\text{idt}}$ simultaneously functions as the transmon qubit shunt capacitor and as the electromechanical transducer (an interdigital transducer, IDT) \cite{gustafsson_propagating_2014} of the SAW resonator, ensuring maximal spatial overlap between the qubit electric field and the mechanical strain field (Fig.~\ref{fig:fig1}b).
Electromechanical transduction arises from the piezoelectric response at the aluminum-silicon interface beneath $C_{\text{idt}}$ (inset of Fig.~\ref{fig:fig1}a).
The qubit is capacitively coupled to a microwave resonator for dispersive readout (RO) and is independently controlled via dedicated XY and flux bias lines. A circuit representation of the entire system is shown in Fig.~\ref{fig:fig1}c, and qubit parameters for the samples studied are listed in Table~\ref{table:S:qubit_params}. These parameters are extracted from the experimental results shown in Fig.~\ref{fig:S:basic}.

The device therefore realizes a qubit coupled to a multimode mechanical resonator (Fig.~\ref{fig:fig1}d), described by the effective Hamiltonian
\begin{equation} \label{eq_hamiltonian}
    \frac{\hat{H}_{\text{qm}}}{\hbar} 
    = \frac{\omega_{\text{q}}}{2} 
            \hat{\sigma}_z
        + \sum_{k=1}^N 
            \left[\omega_{\text{m},k} 
                \hat{b}_k^\dagger \hat{b}_k
        - g_{\text{m},k}
            \hat{\sigma}_x 
            \left( 
                \hat{b}_k + \hat{b}_k^\dagger 
            \right)
            \right],
\end{equation}
where $\omega_{\text{q}} = 2 \pi f_{\text{q}}$ is the qubit transition frequency, $\omega_{\text{m},k}$ denotes the frequency of the $k$th longitudinal SAW resonator mode, $\hat{\sigma}_{x,y,z}$ are the Pauli operators acting in the qubit Hilbert space, and $\hat{b}_k$ ($\hat{b}_k^\dagger$) are the annihilation (creation) operators of the mechanical modes.
The coupling strengths $g_{\text{m},k}$ arise from the interface-mediated electromechanical interaction beneath $C_{\text{idt}}$, whereby strain in the SAW mode induces an electric polarization that couples to the qubit charge degree of freedom.

\section{Detecting interface piezoelectric loss in a transmon qubit}

\begin{figure}
\centering
\includegraphics[width=1.0\columnwidth]{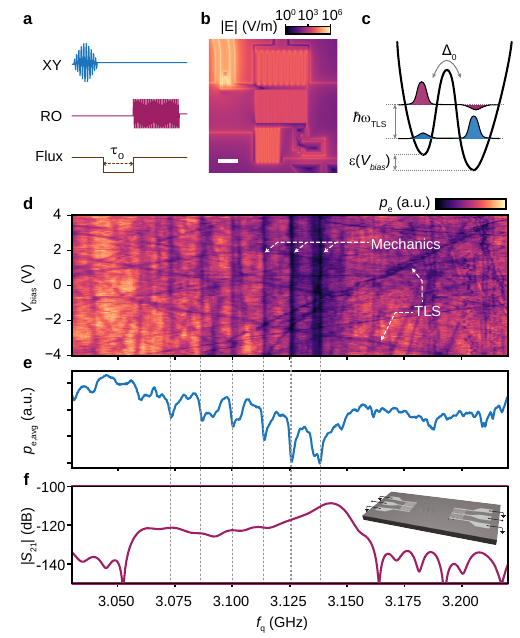}
\vspace{-8mm}
\caption{
\textbf{Detecting interface piezoelectric qubit-phonon coupling.}
\textbf{a}, Pulse sequence used to probe the effect of electromechanical coupling at a fixed delay time $t_{\text{delay}} = \tau_0=3~\mu$s following qubit excitation.
\textbf{b}, Simulated electric-field distribution generated by applying $V_{\text{bias}} = 1~\text{V}$ to the XY line. Scale bar represents 30~$\mu$m.
\textbf{c}, The standard tunneling model of a TLS. An applied voltage bias $V_{\text{bias}}$ shifts the TLS transition energy $\hbar \omega_{\text{TLS}}$.
\textbf{d}, Qubit excited state population $p_e$  measured at delay time $\tau_0$ as a function of qubit frequency $f_{\text{q}}$ and TLS-tuning bias voltage $V_{\text{bias}}$.
\textbf{e}, Bias-averaged $p_e$ at fixed delay time $\tau_0$, revealing the spectral signatures of the SAW resonator modes.
\textbf{f}, Reference transmission through a pair of IDTs with identical dimensions on a separate device shows the operational bandwidth of the transducers (Sample D).
}

\label{fig:fig2}
\end{figure}

Resonant transverse coupling between the qubit and the SAW resonator modifies the dynamics of the excited qubit, leading either to enhanced relaxation in the Purcell regime or to coherent oscillatory exchange in the strong-coupling regime~\cite{blais_cavity_2004, manenti_the_2017, satzinger_quantum_2018}.
The resulting change in qubit dynamics can be monitored by measuring the excited-state population $p_e$ at a fixed delay time $t_{\text{delay}} = \tau_0$ following excitation~\cite{lisenfeld_measuring_2010}, using the pulse sequence shown in Fig.~\ref{fig:fig2}a.
Electromechanical coupling, however, is not the only mechanism capable of modifying $p_e(\tau_0)$. 
First, weakly coupled dielectric TLS defects can enhance energy relaxation and obscure the contribution from piezoelectric coupling. We therefore choose an optimized IDT periodicity, $\lambda_{\text{IDT}} = 1.6 ~\mu\text{m}$, guided by an estimate of the qubit dielectric loss (see Fig. \ref{fig:S:SI_idt_finger_epr}), such that the predicted piezoelectric signature remains resolvable above the TLS-induced loss background.
Second, individual dielectric TLS defects that strongly couple to the qubit can produce qualitatively similar signatures in relaxation measurements~\cite{lisenfeld_measuring_2010}.
It is therefore essential to distinguish collective electromechanical resonances from TLS-induced features.
To this end, we exploit their different responses to externally applied electric fields.
A DC bias voltage $V_{\text{bias}}$ is applied via the qubit XY line (Fig.~\ref{fig:fig1}c), generating an electrostatic field concentrated beneath $C_{\text{idt}}$, as shown in the simulated field distribution in Fig.~\ref{fig:fig2}b for $V_{\text{bias}}=1$~V.
Microscopic TLS defects that interact with the qubit possess electric dipole moments and therefore undergo Stark shifts in an electrostatic field~\cite{grigorij_strain_2012, chen_scalable_2025}, provided that they do not reside in the junctions \cite{Lisenfeld_Efield} and that their dipole moments have sufficient overlap with the applied field.
Within the standard tunneling model~\cite{anderson_anomalous_1972} description, the TLS transition frequency is given by
$
\omega_{\text{TLS}} = \sqrt{\Delta_0^2 + \epsilon^2(V_{\text{bias}})} / \hbar,
$
where $\Delta_0$ is the tunneling amplitude and $\epsilon(V_{\text{bias}})$ is the asymmetry energy that depends on the applied bias voltage (Fig.~\ref{fig:fig2}c) though the dipole interaction.
In contrast, the SAW resonator resonances are set by lithographic geometry and the elastic properties of the host materials, and are therefore largely insensitive to $V_{\text{bias}}$.

Experimentally, we map $p_e(\tau_0)$ as a function of qubit frequency and $V_{\text{bias}}$.
The result (Fig.~\ref{fig:fig2}d) reveals two distinct classes of features.
A set of resonances remains fixed in frequency across the entire bias range, forming vertical bands in the spectrum. These field-independent features are mechanical resonances that coincide with the independently measured IDT passband (Fig.~\ref{fig:fig2}f) and are attributed to qubit-phonon coupling described by the Hamiltonian in Eq.~(\ref{eq_hamiltonian}).
In addition, we observe resonances that shift strongly with $V_{\text{bias}}$, consistent with Stark-tunable TLS defects.
The phonon-induced contribution to $p_e$ can be isolated by averaging the data over $V_{\text{bias}}$, which averages out the effects from TLS resonances while preserving bias-independent mechanical resonances (Fig.~\ref{fig:fig2}e). After averaging over the bias field, a series of dips emerges within the IDT passband (Fig.~\ref{fig:fig2}f), exhibiting nearly uniform spacing corresponding to the free spectral range (FSR), consistent with the discrete longitudinal modes of the reflector-defined SAW resonator.

\section{Interface piezoelectricity-dominated qubit relaxation}

\begin{figure}
\centering
\includegraphics[width=1.0\columnwidth]{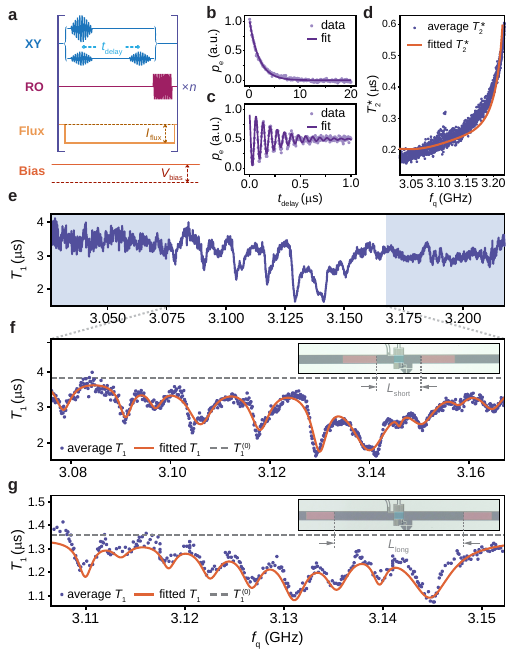}
\vspace{-8mm}
\caption{
    \textbf{
    Qubit lifetime reduction due to piezoelectric coupling to SAW resonator modes.
    }
    \textbf{a}, Pulse sequence used for coherence measurements.
    \textbf{b}, Example of an energy-relaxation ($T_1$) characterization measured at $f_{\text{q}}=$3.14~GHz.
    \textbf{c}, Example of a Ramsey ($T_2^*$) characterization measured at $f_{\text{q}}=$3.14~GHz.
    \textbf{d}, Bias-averaged $T_2^*$ as a function of qubit frequency. The decreasing trend of $T_2^*$ away from the sweet spot is captured by the flux sensitivity $T_2^* \propto (\partial f_{\text{q}} / \partial I_{\text{flux}})^{-1}$ \cite{Koch_transmon}.
    \textbf{e}, Bias-averaged $T_1$ as a function of qubit frequency $f_{\text{q}}$. The white-background region indicates the bandwidth of the SAW mirror.
    \textbf{f}, Zoomed-in view of panel \textbf{e}, fitted using Eq.~(\ref{eq:T1_reduction_model}). The fit yields the background relaxation time $T_1^{(0)}$, indicated by the gray dashed line, as well as the electromechanical coupling strength $g_{\text{m},k}$ and the SAW resonator mode linewidth $\kappa_{\text{m},k}$. (See Table.~\ref{table:extracted_g_m}.) Inset: optical micrograph of Sample A. $L_{\text{short}}=380~\mu\text{m}$ is the distance between the two sets of reflectors.
    \textbf{g}, Same as panel \textbf{f}, but for Sample B, which has a larger resonator length $L_\text{long}$ than Sample A. Inset: optical micrograph of Sample B. $L_{\text{long}} = 1120~\mu$m is the distance between the two sets of reflectors.
}
\label{fig:fig3}
\end{figure}

\begin{figure}
\centering
\includegraphics[width=1.0\columnwidth]{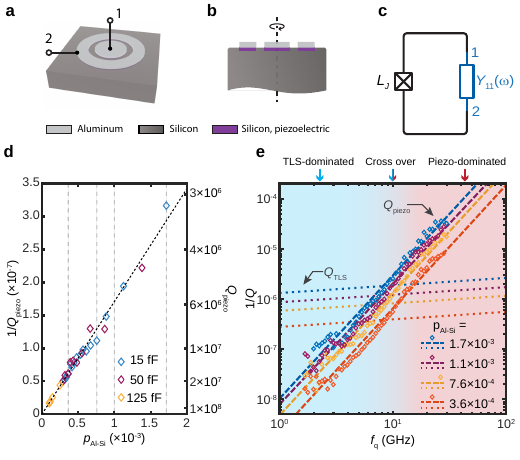}
\caption{
    \textbf{Predicted frequency dependence of interface piezoelectric loss for superconducting qubits.} All quantities are obtained from finite-element simulations of an aluminum-on-silicon coplanar capacitor.
    Schematic \textbf{a-b}, Top (a) and side (b) views of the simulated capacitor. Axial symmetry is applied to reduce computational complexity. Features are not drawn to scale. 
    \textbf{c}, Circuit model of the superconducting qubit studied. $Y_{11}(\omega)$ is the effective admittance of the capacitor, as seen by the Josephson junction $L_{\text{J}}$, including piezoelectric coupling in a multiphysics simulation. 
    \textbf{d}, Simulated EPR-dependent quality factor of the qubit at $f_{\text{q}} = \omega_{\text{q}} / 2\pi = 4.5$~GHz. Diamonds with different colors correspond to capacitors with different dimensions and geometric capacitance.
    \textbf{e}, Simulated frequency dependence of $Q_{\text{piezo}}$ for a 15~fF coplanar capacitor (diamonds). 
    The dashed line is a linear fit on a log-log scale, giving $1/Q_{\text{piezo}} \propto f_{\text{q}}^{2.4}$, or equivalently $\Gamma_{\text{piezo}} = 2\pi f_{\text{q}}/Q_{\text{piezo}} \propto f_{\text{q}}^{3.4}$. The dotted line shows the expected TLS-limited quality factor $Q_{\text{TLS}}$ assuming $\tan \delta_{\text{TLS}} = 10^{-3}$. Colors denote different $p_{\text{Al-Si}}$, marked in panel d by vertical dashed lines.
}
\label{fig:fig4}
\end{figure}

To study the qubit-phonon resonant dynamics in more detail and to extract the electromechanical coupling strengths $g_{\text{m},k}$, we measure the qubit excited-state population as a function of delay time $t_{\text{delay}}$ for different qubit frequencies $\omega_{\text{q}}$. The resulting traces are well fit by single-exponential decays (Fig.~\ref{fig:fig3}b), indicating that the qubit-phonon system is in the weak-coupling regime.
We characterize the frequency dependence of the qubit relaxation rate $\Gamma_{1,\text{q}} = 1/T_{1}$, where $T_1$ is the qubit relaxation time. 
From the Lindblad master equation \cite{barends_Xmon_2013} (also see Supplementary),
\begin{equation}
\label{eq:T1_reduction_model}
    \Gamma_{1,\text{q}}(\omega_{\text{q}})
    = \Gamma_{1,\text{q}}^{(0)}
        + \sum_{k=1}^{N}
            \frac{
                2|g_{\text{m},k}|^2 \Gamma_{2,k}
            }{
                \Gamma_{2,k}^2 + (\omega_{\text{q}} - \omega_{\text{m},k})^2
            } ,
\end{equation}
where the effective qubit-SAW resonator decoherence rates are
$
    \Gamma_{2,k}
    = \Gamma_{2,\text{q}}
        + \kappa_{\text{m},k} / 2 ,
$
and $T_2^* = 1/\Gamma_{2,\text{q}}$ denotes the qubit coherence time. The nominal qubit decay rate $\Gamma_{1,\text{q}}^{(0)} = 1 / T_1^{(0)}$ is added to account for any non-piezoelectric qubit loss. Equation~(\ref{eq:T1_reduction_model}) enables the extraction of both $g_{\text{m},k}$ and $\kappa_{\text{m},k}$ once $\Gamma_{1,\text{q}}$ and $\Gamma_{2,\text{q}}$ are obtained from standard qubit measurements.

Figure~\ref{fig:fig3}a shows the pulse sequences used to measure $T_1$ (top XY sequence) and $T_2^*$ (bottom XY sequence). 
To mitigate excess lifetime fluctuations arising from qubit-TLS coupling, we again employ a bias-averaging protocol: For each combination of TLS bias $V_{\text{bias}}$, SQUID bias $I_{\text{flux}}$, and delay time $t_{\text{delay}}$, the sequence in Fig.~\ref{fig:fig3}a is repeated $n$ times.
At a fixed qubit frequency, determined by $I_{\text{flux}}$, the qubit population as a function of $t_{\text{delay}}$ is extracted by averaging the readout signal over the $n$ repetitions and over $n_{\text{bias}}$ distinct $V_{\text{bias}}$.
Representative bias-averaged energy relaxation and Ramsey decay traces are shown in Fig.~\ref{fig:fig3}b and~\ref{fig:fig3}c, respectively.

Fig.~\ref{fig:fig3}d and \ref{fig:fig3}e show the extracted $T_2^*$ and $T_1$ as functions of the qubit frequency. Pronounced reductions in $T_1$ are observed when the qubit is in resonance with the modes of the SAW resonator, matching the features shown in Fig.~\ref{fig:fig2}d and \ref{fig:fig2}e. The absence of impact on $T_2^*$ indicates the coupling is dominated by the transverse component, consistent with the model in Eq.~\ref{eq_hamiltonian}.
Fitting the $T_1$ spectrum using Eq.~(\ref{eq:T1_reduction_model}) yields coupling strengths $g_{\text{m},k}/2\pi \sim 100~\text{kHz}$ and mechanical linewidths $\kappa_{\text{m},k}/2\pi \sim 1~\text{MHz}$ (see Table \ref{table:extracted_g_m} for more details).

To further validate this interpretation, we performed the same measurements on two additional qubits, Sample B and Sample C (Fig.~\ref{fig:S:images}c–f). 
Sample C has a geometry similar to Sample A, while Sample B is embedded in a SAW resonator with a larger reflector separation, corresponding to a smaller FSR. 
In both cases, we observe similar dips in the $T_1$ spectrum (Fig.~\ref{fig:fig3}g; see also Fig.~\ref{fig:S:pe}), with spacings consistent with the designed FSRs across all three samples (Fig.~\ref{fig:S:pe}).
In addition, we repeated the measurements on Sample C over two cooldown cycles, which reshuffled the TLS distribution, and consistently observed the same mechanical resonances (Fig.~\ref{fig:S:pe_thermalcycle}).
Together, these results rule out the possibility that the biased-averaged resonances originate from electric-field-insensitive TLSs.

The qubit and SAW resonator damping rates exceed the coupling rates, and coherent energy exchange between the qubit and mechanical modes is not resolved in these devices.
Nevertheless, the $T_1$ reduction by up to 50\% unambiguously demonstrates that interface piezoelectricity is a significant dissipation channel for the transmon qubits studied in this work. The observed resonances persist across thermal cycles (Fig.~\ref{fig:S:pe_thermalcycle}) and across three independently fabricated samples (Fig.~\ref{fig:S:pe}).

\section{Energy participation ratio and frequency dependence}

The qubit-phonon coupling mediated by interface piezoelectricity is expected to depend sensitively on device geometry. 
In our devices, the IDT integrated into a SAW resonator enables phase-coherent generation of surface acoustic waves, enhancing electromechanical transduction and, consequently, qubit relaxation. 
By contrast, in conventional planar superconducting qubit geometries, this mechanism is expected to facilitate the incoherent phonon radiation process without the enhancement from the phased-array response of an IDT~\cite{morgan_3_2007}. 
Similar to spontaneous emission from an atomic dipole, the piezoelectric decay rate of a qubit follows Fermi’s golden rule~\cite{jain_acoustic_2023} and is proportional to the product of the interface piezoelectric coupling rate squared (determined by the electric field energy at the interface) and the phonon density of states of the substrate. In the following, we perform finite-element simulations to construct a model that predicts interface piezoelectric qubit dissipation based on a commonly used energy participation ratio analysis~\cite{wang_surface_EPR}. 
To quantify this mechanism in generic device geometries, we perform coupled electrical and mechanical simulations of an aluminum-on-silicon coplanar capacitor (Fig.~\ref{fig:fig4}a,b). Interface piezoelectricity is modeled by assigning a thin piezoelectric layer beneath the electrodes (Fig.~\ref{fig:fig4}b)~\cite{zhou_observation_2025}, and axial symmetry is used to reduce computational complexity. When embedded in a resonant circuit (Fig.~\ref{fig:fig4}c), the capacitor defines a qubit (or resonator) whose quality factor $Q_{\text{piezo}} = \mathrm{Im} Y_{11}(\omega_{\text{q}}) / \mathrm{Re} Y_{11}(\omega_{\text{q}})$ is extracted from the admittance~\cite{supp, zhou_observation_2025}.

We first study the geometric dependence of the simulated interface piezoelectric loss at a fixed qubit frequency $f_{\text{q}} = 4.5~\text{GHz}$. Across the simulated range of coplanar capacitor geometries, the quality factors collapse onto a single linear curve when plotted against the interface energy participation ratio (EPR), $p_{\text{Al-Si}}$, defined as the fraction of electric field energy stored within a 3~nm-thick layer beneath the electrodes~\cite{wang_surface_EPR}, independent of the total capacitance (Fig.~\ref{fig:fig4}d). The linear dependence identifies the EPR as a useful parameter for predicting interface piezoelectric loss. The corresponding loss tangent, $\tan \delta_{\text{piezo}} = 1 / p_{\text{Al-Si}} Q_{\text{piezo}} \approx 1.7 \times 10^{-4}$ at $4.5~\text{GHz}$, lies below the typical TLS loss tangent of state-of-the-art transmons, suggesting that interface piezoelectricity is subdominant in conventional qubit designs at these frequencies.

The picture changes at higher frequencies. For fixed $p_{\text{Al-Si}}$, the simulated piezoelectric decay rate follows a steep power-law $\Gamma_{\text{piezo}} \propto f_{\text{q}}^{3.4}$ across the frequency range studied (Fig.~\ref{fig:fig4}e), consistent with a Fermi’s golden rule estimate based on the phonon density of states~\cite{geller_local_2004} and zero-point fluctuations. The exponent is approximately independent of EPR, suggesting that geometry (through EPR) sets the coupling strength at a given frequency, whereas the frequency dependence is governed by the phononic environment.
In contrast, dielectric loss is only weakly frequency dependent and can be captured by a phenomenological model~\cite{nguyen_high_2019, supp}. Combining these trends, our simulations predict that interface piezoelectric loss becomes comparable to TLS loss near $10$~GHz and dominates at higher frequencies. The precise crossover frequency depends on the substrate and heterostructure, but the steep frequency scaling is set by the bulk phonon density of states. 

For clarity, we summarize the properties of the interface piezoelectric loss and other well-recognized loss channels~\cite{mcrae_materials_2020} in Table~\ref{table:S:loss_channels}.

\section{Discussion}
Interface piezoelectricity presents a distinct challenge for superconducting qubits. Unlike conventional loss mechanisms attributed to material disorder, it originates from broken symmetry at the metal-substrate interface and is therefore a structurally rooted decoherence channel that is not expected to be eliminated by improving material quality alone. 
Our simulations indicate that interface piezoelectricity does not limit state-of-the-art qubits operating in the sub-10~GHz regime; however, continued reductions in the TLS loss tangent may soon bring it into focus even below 10~GHz.

The interface piezoelectric loss poses a challenge to superconducting qubits operating at high frequencies, such as those in the millimeter-wave regime.
Millimeter-wave qubits are attracting growing interest because of their unique advantages, such as relaxed requirements on environmental temperature, compact component size, and faster gate times~\cite{anferov_millimeter_2025}. The steep frequency scaling predicted by our simulations suggests that interface piezoelectric loss may become a key obstacle in this regime. 

Mitigating this structurally rooted dissipation mechanism will require device-level engineering, rather than materials optimization alone.
One route is to reshape the phonon density of states using quasi-two-dimensional membrane geometries, and thereby modify the frequency scaling of the loss (Fig.~\ref{fig:S:bulk_membrane}).
Building on such membrane-based platforms, phononic metamaterials offer a more targeted and designable approach: their engineered band structures can be tailored to remove available phonon modes within the frequency range of interest~\cite{odeh_non-markovian_2025, chen_phonon_2024}.
Alternatively, engineering the spatial profiles of the microwave and acoustic modes to minimize field overlap at the interface can suppress the coupling hotspot (see undercut engineering in Fig.~\ref{fig:S:undercuts}).

On the other hand, the same mechanism can potentially serve as a flexible and efficient coupling mechanism for quantum manipulation of phonons. 
Efficient coupling between electromagnetic and mechanical degrees of freedom remains a central challenge in hybrid quantum acoustic systems~\cite{chu_quantum_2017}.
In our platform, appropriate co-design of the qubit and acoustic resonator could enable coherent energy exchange mediated by interface piezoelectricity, with applications to quantum acoustic devices.

\section{Methods}
\subsubsection{Sample fabrication}
Samples A-D were fabricated by patterning superconducting aluminum on an undoped silicon substrate, combining etch and lift-off processes.
First, a 50~nm aluminum thin film was deposited onto a 100~mm undoped silicon wafer (resistivity $\geq$ 20,000~$\Omega \cdot {\text{cm}}$) using an electron-beam evaporator. The wafer was first cleaned with RCA-1, followed by a buffered oxide etch. The sample was then immediately transferred to the load lock of the electron-beam evaporator.
The wafer was subsequently diced into 10~mm-by-10~mm pieces for further processing.
For Sample A and Sample B, the IDT and Bragg reflectors were patterned using electron beam lithography, followed by plasma etch of aluminum.
Then, the coplanar waveguides and ground plane were formed by photolithography, followed by a wet etch.
After the wet etch, the Manhattan-style Josephson junctions were defined using electron-beam lithography. Prior to aluminum deposition by shadow-mask evaporation in an electron-beam evaporator, the samples were cleaned by an oxygen plasma treatment, followed by a dip in 6\% buffered oxide etch solution.
Finally, a bandage layer was defined by electron-beam lithography. To form galvanic contact, the exposed surfaces were cleaned by in-situ argon ion milling, followed by aluminum deposition in the same electron-beam evaporator.
Sample C was fabricated using a process similar to those used for Samples A and C.
The only difference is that the IDTs, Bragg reflectors, coplanar waveguides, and ground plane were patterned together in a single electron-beam lithography step, followed by a plasma etch.
Sample D was fabricated by first defining the IDTs using electron-beam lithography and plasma etch, and then defining the rest of the features using photolithography and a wet etch.

\subsubsection{Measurement}
All measurements were carried out in a dilution refrigerator at the base temperature of about 7~mK.
The chip was wire-bonded to a printed circuit board, mounted in a copper sample box, and thermalized to the mixing-chamber flange of the dilution refrigerator. 
For qubit measurement, the samples were enclosed in $\mu$-metal shields to suppress magnetic noise. 
Control and readout pulses were generated using a Quantum Machine OPX+ system and upconverted at room temperature using IQ mixers and local oscillators. 
A low-noise DC source provided the bias voltage, which was combined with the microwave drive using a bias tee at the mixing-chamber flange. 
The input lines were attenuated and filtered at multiple temperature stages to suppress thermal noise. 
The reflected readout signal was routed through circulators for isolation, amplified by a high-electron-mobility transistor amplifier at 4~K and by additional low-noise amplifiers at room temperature, then downconverted to an intermediate frequency and digitized by the OPX+ for quadrature demodulation.
The detailed measurement setup for the qubit-SAW resonator system is shown in Fig.~\ref{fig:S:instrument}.
The reference transmission coefficient of the delay-line IDT (Sample D) was obtained in the same system using a vector network analyzer.

\subsubsection{Numerical Simulation}
Data shown in Fig.~\ref{fig:S:SI_idt_finger_epr} were obtained using Ansys Electronics 2024 R1 running on Microsoft Windows 11 Pro for Workstations. All other finite-element simulation results were obtained using COMSOL Multiphysics 6.0 running on Microsoft Windows 11 Pro for Workstations.

\section{Data availability}
All experimental data and COMSOL simulation model files that support this study are available at https://doi.org/10.5281/zenodo.20150434.

\section{Acknowledgments}
This work was primarily funded by the U.S. Department of Energy, Office of Science, Office of Basic Energy Sciences, Materials Sciences and Engineering Division under Contract No. DE-AC02-05CH11231 within the Quantum Coherent Systems Program (KCAS26). Additional support was
provided by the Air Force Office of Scientific Research and the
Office of Naval Research under Grant No. FA9550-23-1-
0333 (lithography development), by the U.S. Department of Energy, Office of Science, Basic Energy Sciences, Materials Sciences and Engineering Division under Contract No. DE-AC02-05CH11231 within the Phonon Control for Next-Generation Superconducting Systems and Sensors program KCAS23 (material optimization). A.S. acknowledges support from the Society of Hellman Fellows. The devices used in this work were fabricated at Berkeley Lab's Molecular Foundry and UC Berkeley's Marvell Nanofabrication Laboratory.

\section{Author Contributions}
A. Sipahigil conceived and supervised the work. H.Z. and K.Y. designed the devices. H.Z., K.Y., and Y.B. fabricated the devices with assistance from L.S., Z.-H.Z., and A. Schwartzberg. K.Y., H.Z., and Z.-H.Z. built the measurement setup. H.Z. and K.Y. performed the measurements and analyzed the data. K.Y., S.S., H.Z., and A. Sipahigil performed the theoretical modeling and simulations. H.Z., K.Y., and A. Sipahigil wrote the manuscript with input from all authors.

\section{Competing interests}
The authors declare no competing interests.

\normalem
\let\oldaddcontentsline\addcontentsline
\renewcommand{\addcontentsline}[3]{}

\let\addcontentsline\oldaddcontentsline

\clearpage
\pagebreak
\onecolumngrid

\setcounter{equation}{0}
\setcounter{table}{0}
\setcounter{figure}{0}
\setcounter{page}{1}
\setcounter{section}{0}
\makeatletter
\renewcommand{\theequation}{S\arabic{equation}}
\renewcommand{\thefigure}{S\arabic{figure}}
\renewcommand{\thetable}{S\arabic{table}}
\renewcommand{\thepage}{S\arabic{page}}

\clearpage
\pagebreak
\onecolumngrid
\begin{center}
\textbf{\large Supplementary Information for ``Interface Piezoelectric Loss in Superconducting Qubits'' }\\[5pt]
\end{center}

\begin{table*}[h]
\begin{center}
   \caption{Qubit parameters for the samples studied. A large qubit shunt capacitor is used for two reasons. First, the piezoelectric coupling increases with the number of IDT periods. Second, longer IDT fingers reduce acoustic diffraction, thereby increasing the quality factor of the SAW cavity.}
    \label{table:S:qubit_params}
    \begin{tabular}{|c||c|c|c|c|c|c|}
    \hline 
     & \# IDT periods & IDT periodicity ($\mu$m) & IDT finger length ($\mu$m) & $E_C/h$ (MHz) & $E_{\text{J,max}}/h$ (GHz) & $f_{\text{q,max}}$ (MHz)\\ \hline
    Sample A & 50 & 1.6 & 48 & 68 & 20.0 & 3233\\
    Sample B & 50 & 1.6 & 48 & 73 & 20.5 & 3388\\
    Sample C & 50 & 1.6 & 48 & 67 & 21.7 & 3344\\ \hline
    \end{tabular}
\end{center}
\end{table*}

\begin{table}[h!]
\setlength{\tabcolsep}{6pt}
\centering
\caption{Extracted electromechanical coupling strength $g_k$ and mechanical linewidth $\kappa_{\text{m},k}$ for Samples A, B, and C from the fitting of Eq.~(\ref{eq:T1_reduction_model}).}
\begin{tabular}{|c||c|c|c||c|c|c|}
\hline
\multirow{2}{*}{Resonance $k$}
 & \multicolumn{3}{c||}{\makecell[c]{Coupling $g_k / 2\pi$ \\(kHz)}} 
 & \multicolumn{3}{c|}{\makecell[c]{SAW resonator\\linewidth $\kappa_{\text{m},k} / 2\pi$ \\ (MHz)}} \\
\cline{2-7}
 & Sample A & Sample B & Sample C 
 & Sample A & Sample B & Sample C \\
\hline
$1$ & 100 & 78 & 50 & 2.25 & 0.66 & 1.52 \\
$2$ & 101 & 52 & 120& 1.61 & 1.03 & 4.22 \\
$3$ & 83 & 63 & 139 & 2.00 & 0.70 & 2.16 \\
$4$ & 159 & 87 & 166 & 4.03 & 1.27 & 2.21 \\
$5$ & 149 & 99 & 259 & 2.78 & 1.34 & 6.86 \\
$6$ & 174 & 128 & 113 & 1.67 & 1.80 & 6.13 \\
$7$ & 272 & 109 & 90 & 5.53 & 1.76 & 3.81 \\
$8$ & 45 & 72 & - & 0.10 & 0.66 & - \\
$9$ & 154 & 168 & - & 4.80 & 3.52 & - \\
$10$ & 66 & - & - & 2.37 & - & - \\
$11$ & 115 & - & - & 4.00 & - & - \\
\hline
\end{tabular}
\label{table:extracted_g_m}
\end{table}

\begin{table*}[h]
\begin{center}
    \caption{Comparison of interface piezoelectric loss with other superconducting qubit loss channels. Here, ``L'' and ``C'' indicate whether the loss is associated with inductive or capacitive components, respectively. Columns 3--6 indicate positive (+), negative (–), absent (No), or unknown (blank) correlations between each loss channel and the corresponding experimental parameter. ``Yes'' indicates that a correlation exists. qp-IR denotes quasiparticle loss due to stray infrared light, and qp-$\mu$w denotes quasiparticle loss due to microwave-induced pair breaking. ``Dimension'' refers to the conductor/gap widths of coplanar waveguides and IDT capacitors for TLS and radiation loss, and to the inductor conductor width for vortex loss. Except for the last row, the results, including loss-channel definitions and correlation notation, are adapted from Ref.~\cite{mcrae_materials_2020}.
    }
    \label{table:S:loss_channels}
    \begin{tabular}{|c||c|c|c|c|c|}
    \hline
    Loss Type & L or C & Power & Temperature & Frequency & Dimension\\
    \hline
    TLS & C & – & – & + & – \\
    qp-Thermal & L & No & + & + & \ \\
    qp-IR & L & No & No & \ & \ \\
    qp-$\mu$w & L & + & + & + & \ \\
    vortex & L & No & + & + & + \\
    radiation & L, C & No & No & + & + \\
    parasitic modes & L, C & No & No & Yes & \ \\
    \textbf{interface piezo} & \textbf{C} & \textbf{No} & \textbf{No} & \textbf{+} & \textbf{–} \\
    \hline
    \end{tabular}
\end{center}
\end{table*}

\begin{figure*}[h!]
\centering
\includegraphics[width=\textwidth]{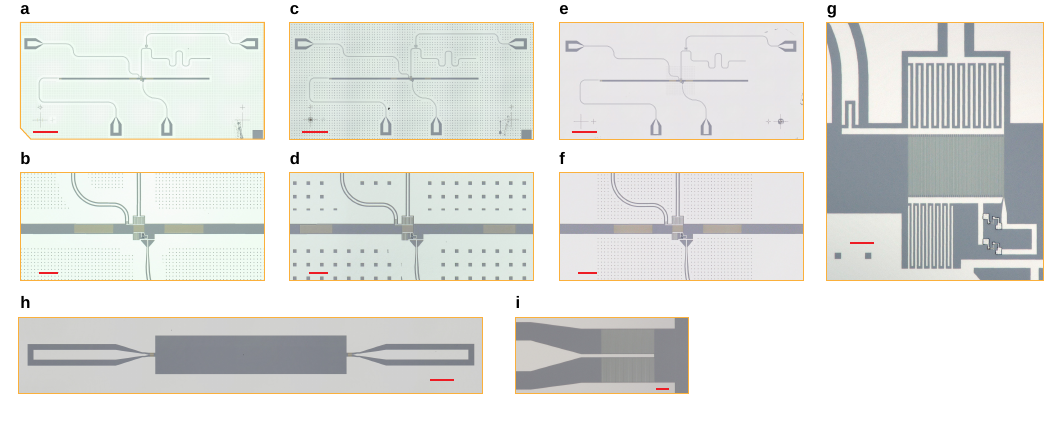}
\caption{\textbf{Optical micrographs of the devices studied.}
\textbf{a}, Optical micrograph of Sample A. The scale bar represents 1~mm.
\textbf{b}, Optical micrograph of Sample A, zoomed in on the transmon-cavity system. The scale bar represents 200~$\mu$m.
\textbf{c-d}, Same as panel a-b, for Sample B.
\textbf{e-f}, Same as panel a-d, for Sample C.
\textbf{g}, Optical micrograph of Sample C, zoomed in on the transmon region. The scale bar represents 25~$\mu$m.
\textbf{h}, Optical micrograph of Sample D. Scale bar represents 500~$\mu$m.
\textbf{i}, Optical micrograph of Sample D, zoomed in on the left transducer. Scale bar represents 20~$\mu$m.
}
\label{fig:S:images}
\end{figure*}

\begin{figure*}[!h]
\centering
\includegraphics[width=0.8\textwidth]{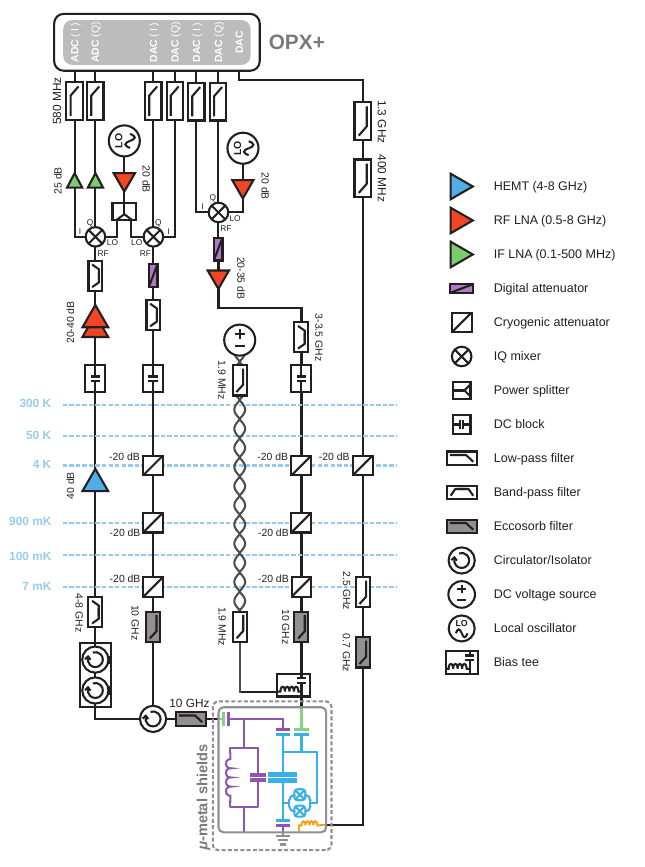}
\caption{\textbf{Measurement setup.} Schematic of the cryogenic microwave circuit used for qubit control and readout. Microwave signals are generated and upconverted at room temperature, attenuated and filtered before reaching the device at base temperature, and the output is isolated, amplified, downconverted, and digitized for measurement. HEMT: high-electron-mobility-transistor amplifier. RF/IF LNA: radio-frequency/intermediate-frequency low-noise amplifier.}
\label{fig:S:instrument}
\end{figure*}

\pagebreak

\begin{figure*}[!h]
\centering
\includegraphics[width=\textwidth]{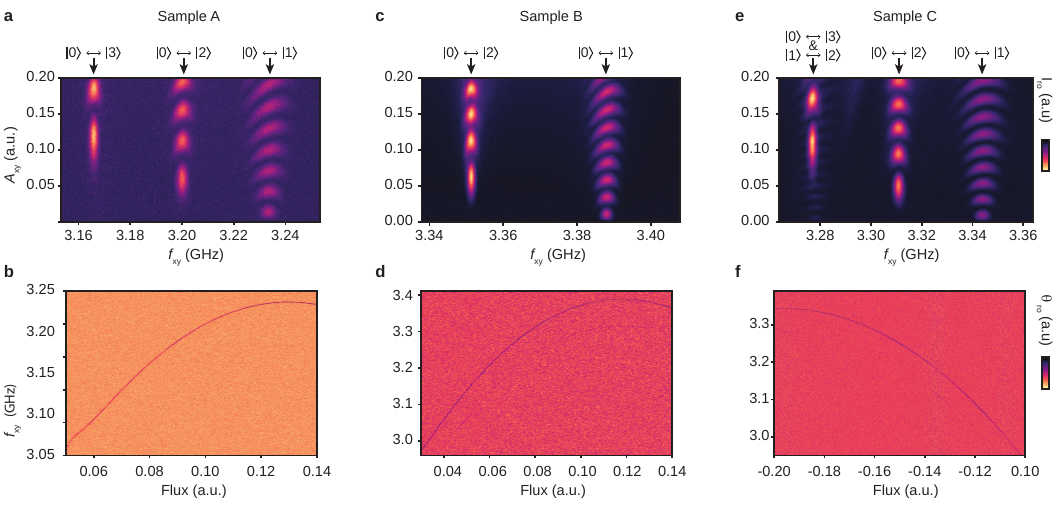}
\caption{\textbf{Rabi oscillations and qubit spectra.}
\textbf{a} Rabi oscillation of Sample A as a function of the XY drive frequency $f_{\text{xy}}$ and drive amplitude $A_{\text{xy}}$ measured at the maximal qubit frequency. $I_{\text{ro}}$ is the readout amplitude.
\textbf{b} Two-tone spectrum of Sample A as a function of flux and XY drive frequency $f_{\text{xy}}$. $\theta_{\text{ro}}$ is the readout phase.
\textbf{c-d} Same as a-b, for Sample B.
\textbf{e-f} Same as a-d, for Sample C.
}\label{fig:S:basic}
\end{figure*}

\begin{figure*}[h!]
\centering
\includegraphics[width=\textwidth]{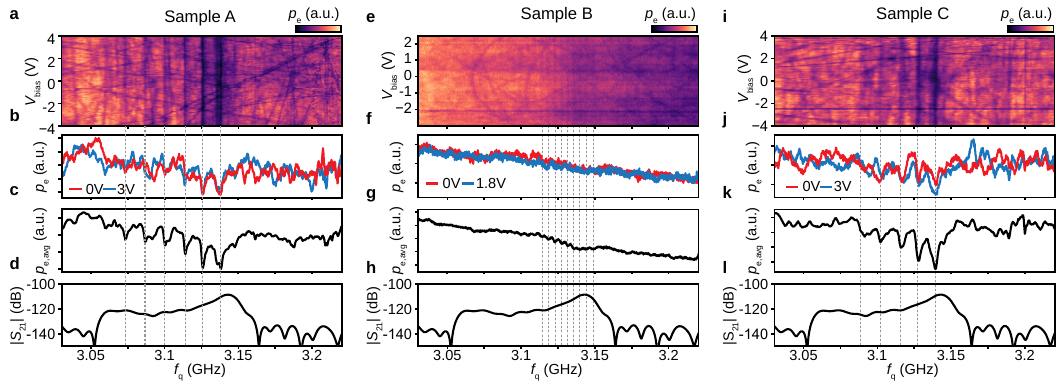}
\caption{\textbf{Electric field dependent population measurement on different samples}
\textbf{a}, Qubit excited state population $p_e$  measured at delay time $\tau_0=3~\mu$s as a function of qubit frequency $f_{\text{q}}$ and bias voltage $V_{\text{bias}}$. The same plot is also shown in Fig.~\ref{fig:fig2}d.
\textbf{b}, Line-cuts of panel a at fixed bias voltage.
\textbf{c}, Bias-averaged $p_e$ at fixed delay time $\tau_0=3~\mu$s, revealing the spectral signatures of the SAW resonator modes. The same plot is also shown in Fig.~\ref{fig:fig2}e.
\textbf{d}, Transmission coefficient as a function of frequency measured in Sample D, the reference transducer. The same plot is also shown in Fig.~\ref{fig:fig2}f.
\textbf{e-h}, Same as panel a-d, for Sample C.
\textbf{i-l}, Same as panel a-h, for Sample B. The larger size of the SAW cavity results in a smaller FSR, and therefore, more modes are resolved within the passband of the IDT. Yet the qubit-SAW resonator coupling rate $g_{\text{m},k}$ is also weaker, since, approximately, $g_{\text{m},k} \propto \sqrt{\text{FSR}}$.
}\label{fig:S:pe}
\end{figure*}

\begin{figure*}[h!]
\centering
\includegraphics[width=\textwidth]{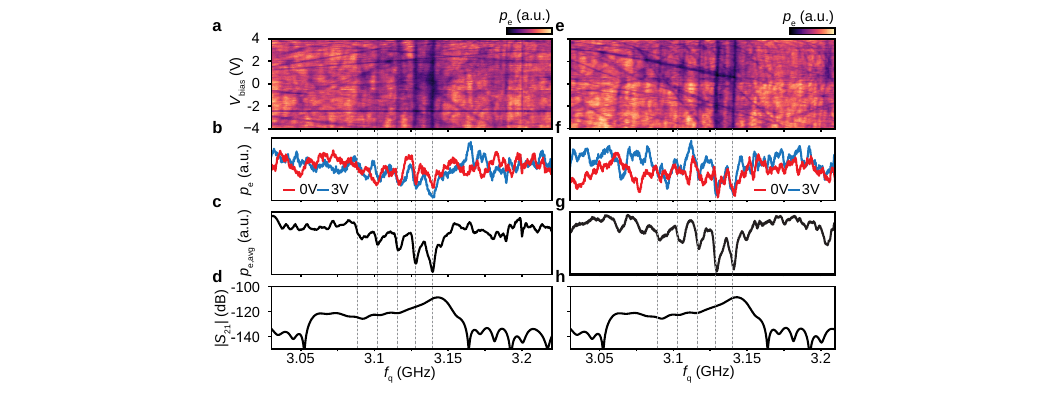}
\caption{\textbf{Measurements of Sample C in different cooldowns.}
\textbf{a-d}, Same as Fig.~\ref{fig:S:pe}e-h, plotted in a smaller range of $f_{\text{q}}$. Data were taken at a delay time $\tau_0=3~\mu$s.
\textbf{e-h}, Same as panel a-d. Data are collected in a different thermal cycle. Note that the spectrum of the SAW resonator is consistent, while the TLS spectrum changed significantly.
The small shift of the SAW modes is due to the slow drift of the qubit frequency during the experiment.  Data were taken at a delay time $\tau_0=3~\mu$s.
}\label{fig:S:pe_thermalcycle}
\end{figure*}

\begin{figure*}[!h]
\centering
\includegraphics[width=1\textwidth]{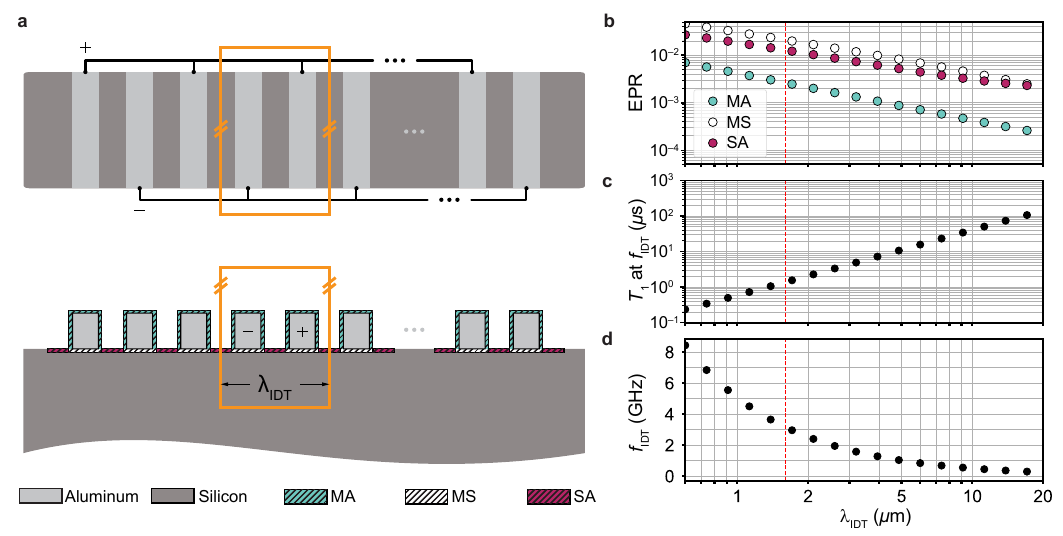}
\vspace{-4mm}
\caption{\textbf{Surface participation ratio and dielectric loss of the IDT qubits.} 
\textbf{a,} Schematic of the electrostatic model used to extract the interface energy participation ratio (EPR) of an aluminum-on-silicon IDT structure. The top and bottom drawings show the top and side views, respectively. Since the finger length, $48~\mu\text{m}$, exceeds the acoustic wavelengths $\lambda_{\text{IDT}}$ of interest, the electric field is assumed invariant along the fingers, reducing the problem to two dimensions. The actual IDT qubits have 50 periods, allowing the electric field to be further approximated as periodic; therefore, only a single period, highlighted by the orange boxes, is simulated with periodic boundary conditions along the propagation direction of SAW. Interface energy is calculated by introducing a $3$-nm dielectric layer with a dielectric constant of $10$ at the metal–air (MA), metal–substrate (MS), and substrate–air (SA) interfaces. 
\textbf{b,} Surface participation ratio versus IDT periodicity (i.e., acoustic wavelength), with separate contributions from MA, MS, and SA interfaces.
\textbf{c,} Estimated TLS-limited qubit lifetime as a function of IDT periodicity, evaluated with the qubit flux-biased to the center of the IDT passband. A loss tangent of $\tan\delta_{\text{TLS}} = 10^{-3}$ is assumed.
\textbf{d,} Center frequency of the IDT passband versus IDT periodicity. The red dashed line in each panel marks the parameters of our devices ($\lambda_{\text{IDT}} = 1.6~\mu\text{m}$). 
}\label{fig:S:SI_idt_finger_epr}
\end{figure*}

\begin{figure*}[!h]
\centering
\includegraphics[width=\textwidth]{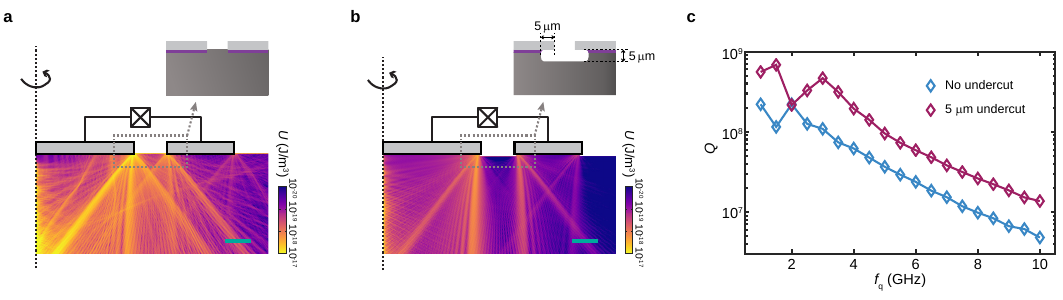}
\vspace{-8mm}
\caption{\textbf{Reducing interface piezoelectric loss with hotspot engineering. }
Simulations using finite-element methods indicate that the interface piezoelectricity-induced acoustic radiation is most significant below the edges of the superconductor electrodes, since there the strong electric fields overlay with the piezoelectric interfaces. Removing the silicon beneath the aluminum edges will both alter the electric field distribution and eliminate the interface piezoelectricity in these regions.
\textbf{a}, Mechanical energy density of a qubit with a coplanar shunt capacitor. The high-energy regions are shown as ``jet''-shaped features that originate from the edges of the aluminum electrodes. Scale bar represents 50~$\mu$m.
\textbf{b}, Same as a, for capacitors with an undercut profile beneath the boundaries of the aluminum electrodes. The overall energy density is lower, and the phonon injection from the electrode edges is significantly suppressed. The profile of the undercut is schematically shown as an inset. Scale bar represents 50~$\mu$m.
\textbf{c}, Interface piezoelectric loss-limited qubit quality factor $Q$ as a function of qubit frequency $f_{\text{q}}$ for qubits with and without the edge undercuts. Undercut leads to an improved $Q$ as shown from $f_{\text{q}} = 1$~GHz to 10~GHz.
}\label{fig:S:undercuts}
\end{figure*}

\begin{figure*}[!h]
\centering
\includegraphics[width=\textwidth]{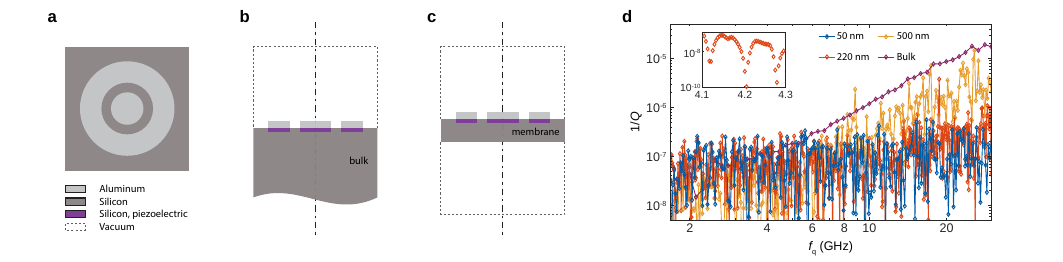}
\caption{\textbf{Reducing high-frequency interface piezoelectric loss with superconducting qubits on silicon membranes.}
In the main text, we show that interface piezoelectric loss in superconducting qubits becomes more pronounced at high frequencies. One possible mechanism is the acoustic phonon density of states, which scales as $\rho \sim f_{\text{q}}^2$ in three-dimensional substrates, but only as $\rho \sim f_{\text{q}}$ in two-dimensional systems. This suggests that fabricating high-frequency qubits on quasi-two-dimensional substrates may help suppress interface piezoelectric loss.
To explore this possibility, we simulated the interface piezoelectric loss of superconducting qubits with coplanar capacitors on silicon membranes of different thicknesses using the same finite-element framework developed for bulk-substrate simulations.
\textbf{a}, Schematic top view of the coplanar capacitor on both bulk and membrane substrates.
\textbf{b}, Schematic side view of the coplanar capacitor on a bulk silicon substrate, corresponding to the structure studied in Fig.~\ref{fig:fig4} of the main text.
\textbf{c}, Schematic cross-sectional side view of the coplanar capacitor on a silicon membrane.
\textbf{d}, Inverse quality factor $1/Q$ as a function of qubit frequency $f_{\text{q}}$ for the simulated structures. Colors indicate different substrates, including bulk silicon and silicon membranes of varying thicknesses. The aluminum thickness is 50~nm for the bulk substrate and for membranes with thicknesses of 220~nm and 500~nm, and 10~nm for the 50~nm-thick membrane. Inset: finer frequency sweep near $f_{\text{q}} = 4.2$~GHz for the 220~nm-thick membrane.
In these simulations, the lateral capacitor geometry is kept fixed across all substrates, resulting in different capacitance values due to the modified dielectric environment. The noisy features are likely associated with mechanical resonances of the membrane structure, as illustrated more clearly in the inset. Compared to bulk substrates, the interface piezoelectric loss on membrane substrates exhibits a weaker frequency dependence. However, the detailed relationship between $1/Q$ and $f_{\text{q}}$ depends on membrane thickness and does not follow the simple scaling expected from the Fermi's golden rule model. Further investigation is needed to understand these results and to evaluate the potential of membrane-based structures for suppressing interface piezoelectric loss in high-frequency superconducting qubits.}
\label{fig:S:bulk_membrane}
\end{figure*}

\clearpage
\pagebreak

\section{The Circuit QED Model of Electromechanical Coupling}
\begin{figure*}[h!]
\centering
\includegraphics[width=\textwidth]{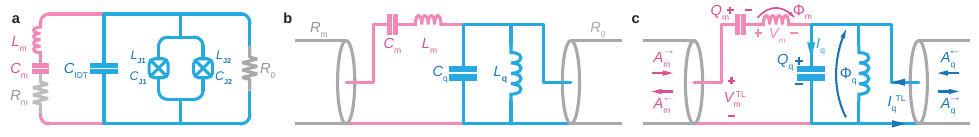}
\caption{
    \textbf{a}, Classical equivalent circuit for a qubit coupled to a SAW cavity mode. The mechanical mode is modeled by a single-mode BVD circuit. 
    \textbf{b}, Linearized circuit with each dissipative element replaced by a transmission line. $C_{\text{q}} = C_{\text{idt}} + C_{\text{J1}} + C_{\text{J2}} \approx C_{\text{idt}}$ is the total capacitance of the qubit and $L_{\text{q}}$ is the effective Josephson inductance of the SQUID at a given flux bias.
    \textbf{c,} Annotated circuit for deriving the equations of motion. The four system observables, $\Phi_{\text{q}}, Q_{\text{q}}, Q_{\text{m}}$, and $\Phi_{\text{m}}$, are coupled to the environment through traveling waves $A_{\text{q}}^{\leftrightarrows}$ and $A_{\text{m}}^{\leftrightarrows}$.
}\label{fig:SI_fig2}
\end{figure*}

In this section, we develop a quantum model for a qubit coupled to a discrete set of mechanical modes. This framework enables the calculation of the electromechanical coupling strengths $g_{\text{m},k}$ once the mechanical modes are represented by an effective electrical circuit. To illustrate the essential ideas with minimal notation, we begin with the simplest case of a qubit coupled to a single mechanical mode; the generalization to multiple modes is straightforward.

We begin from the classical Butterworth-Van Dyke (BVD) description of a SAW resonator, in which the IDT capacitance $C_{\text{idt}}$ is connected in parallel with a motional LC branch \cite{datta_surface_1986}. The qubit circuit is then obtained by identifying $C_{\text{idt}}$ as the qubit shunt capacitance and introducing an additional inductive element to represent the SQUID. The total capacitance of the qubit is denoted by $C_{\text{q}} = C_{\text{idt}} + C_{\text{J}1} + C_{\text{J}2} \approx C_{\text{idt}}$, and the effective junction inductance at a given flux bias is $L_{\text{q}}$. Dissipation in the qubit and mechanical subsystems is incorporated phenomenologically through resistors $R_0$ and $R_{\text{m}}$, respectively. Within this description, the mechanical mode is represented by a series RLC circuit, whereas the qubit mode is represented by a parallel RLC circuit (Fig.~\ref{fig:SI_fig2}a).

Because explicit resistive elements are incompatible with unitary quantum evolution, we follow the standard procedure of replacing each resistor by a lossless transmission line whose characteristic impedance is equal to the corresponding resistance \cite{vool_devoret_cQED_intro}. We denote the transmission line associated with mechanical dissipation by ``M'' and that associated with the nominal qubit dissipation by ``Q''. The resulting lossless circuit is shown in Fig.~\ref{fig:SI_fig2}b.

For the parallel resonator, we take $(\Phi_{\text{q}}, Q_{\text{q}})$ as the conjugate variables, whereas for the series resonator we use $(Q_{\text{m}}, \Phi_{\text{m}})$. Consequently, the canonical commutation relations (or classically the Poisson brackets) of the two modes differ by a minus sign. In addition, the system is driven by two input fields, $A_{\text{q}}^{\leftarrow}$ and $A_{\text{m}}^{\rightarrow}$, which also generate the corresponding reflected output fields. The annotated circuit is shown in Fig.~\ref{fig:SI_fig2}c.

\subsubsection{Classical equations of motion} 
Let $I_{\mu}^{\text{TL}}$ be the total current going from the transmission line $\mu \in \{\text{Q}, \text{M}\}$ into the system and let $V_{\mu}^{\text{TL}}$ be the total voltage at the terminal of the transmission line $\mu$. The total current and voltage on each transmission line can be decomposed into the left and right traveling currents ($I_{\text{q}}^{\leftrightarrows}$, $I_{\text{m}}^{\leftrightarrows}$) and voltages ($V_{\text{q}}^{\leftrightarrows}$, $V_{\text{m}}^{\leftrightarrows}$):
\begin{gather}
    V_{\text{q}}^{\text{TL }}
    = V_{\text{q}}^{\leftarrow } + V_{\text{q}}^{\rightarrow },
    \quad 
    I_{\text{q}}^{\text{TL }}
    = I_{\text{q}}^{\leftarrow } - I_{\text{q}}^{\rightarrow },
\\
    V_{\text{m}}^{\text{TL }}
    = V_{\text{m}}^{\leftarrow } + V_{\text{m}}^{\rightarrow },
    \quad 
    I_{\text{m}}^{\text{TL }}
    = I_{\text{m}}^{\rightarrow } - I_{\text{m}}^{\leftarrow }.
\end{gather}
In addition, classical microwave theory \cite{pozar_microwave} allows us to relate the traveling currents to the traveling voltages using the line impedances:
\begin{equation}
    A_{\mu}^{\leftrightarrows} 
    \triangleq \sqrt{R_{\mu}} I_{\mu}^{\leftrightarrows}
    = \frac{V_{\mu}^{\leftrightarrows}}{\sqrt{R_{\mu}}}, 
    \quad 
    \mu \in \{\text{q}, \text{m}\},
\end{equation}
where $A_{\mu}^{\leftrightarrows}$ are normalized such that $|A_{\mu}^{\leftrightarrows}|^2$ has the unit of power. Applying Kirchhoff's current and voltage laws (i.e., Hamilton’s equations for the generalized momenta) yields
\begin{align} \label{eq:KCL_for_TL_Q}
    I_{\text{q}}^{\text{TL }}
    &= I_{\text{q}} + \frac{\Phi_{\text{q}}}{L_{\text{q}}} - \dot{Q}_{\text{m}}, 
\\ \label{eq:KVL_for_TL_M}
    V_{\text{m}}^{\text{TL}}
    &= \dot{\Phi}_{\text{q}} + \frac{Q_{\text{m}}}{C_{\text{m}}} + V_{\text{m}},
\end{align}
where we have defined $I_{\text{q}} = \dot{Q}_{\text{q}}$ and $V_{\text{m}} = \dot{\Phi}_{\text{m}}$.
There are two other trivial equations (corresponding to Hamilton’s equations for the generalized coordinates):
\begin{align}
    \dot{\Phi}_{\text{q}}
    = \frac{Q_{\text{q}}}{C_{\text{q}}} ,
\qquad
    \dot{Q}_{\text{m}}
    = \frac{\Phi_{\text{m}}}{L_{\text{m}}} .
\end{align}

We do not want to have both the incoming and outgoing waves in the equations, as they are related by the boundary reflection. To remove $I_{\text{q}}^{\rightarrow }$ for the transmission line Q, we use the boundary condition
\begin{equation}
    I_{\text{q}}^{\text {TL}}
    = \underbrace{\frac{A_{\text{q}}^{\leftarrow }}{\sqrt{R_{\text{q}}}}}_{I_{\text{q}}^{\leftarrow }}
        - \underbrace{\frac{1}{R_{\text{q}}}\left(V_{\text{q}}^{\text {TL}}-V_{\text{q}}^{\leftarrow }\right)}_{I_{\text{q}}^{\rightarrow }}
    = \frac{A_{\text{q}}^{\leftarrow }}{\sqrt{R_{\text{q}}}}- \frac{1}{R_{\text{q}}}\left(\dot{\Phi}_{\text{q}} - \sqrt{R_{\text{q}}} A_{\text{q}}^{\leftarrow }\right)
    = 2 \frac{A_{\text{q}}^{\leftarrow }}{\sqrt{R_{\text{q}}}} - \frac{\dot{\Phi}_{\text{q}}}{R_{\text{q}}}.
\end{equation}
Similarly, to remove $V_{\text{m}}^{\leftarrow }$ for the transmission line M, we use
\begin{equation}
    V_{\text{m}}^{\text{TL}}
    =\underbrace{\sqrt{R_{\text{m}}} A_{\text{m}}^{\rightarrow }}_{V_{\text{m}}^{\rightarrow }}
    + \underbrace{R_{\text{m}}\left(-I^{\text{TL }}_{\text{m}} + I_{\text{m}}^{\rightarrow }\right)}_{V_{\text{m}}^{\leftarrow }}
    = \sqrt{R_{\text{m}}} A_{\text{m}}^{\rightarrow }
    + R_{\text{m}}\left(- \dot{Q}_{\text{m}} + \frac{A_{\text{m}}^{\rightarrow }}{\sqrt{R_{\text{m}}}}\right)
    = 2 \sqrt{R_{\text{m}}} A_{\text{m}}^{\rightarrow }
    - R_{\text{m}} \dot{Q}_{\text{m}}.
\end{equation}
Substituting these boundary conditions into Eq.~(\ref{eq:KCL_for_TL_Q}) and (\ref{eq:KVL_for_TL_M}) yields the full set of equations of motion:
\begin{align} \label{eq:classical_EOM_Phi_q}
    \dot{\Phi}_{\text{q}}
    &=\frac{Q_{\text{q}}}{C_{\text{q}}} , 
\\ \label{eq:classical_EOM_Q_q}
    \dot{Q}_{\text{q}} 
    &= I_{\text{q}}=\frac{\Phi_{\text{m}}}{L_{\text{m}}}-\frac{\Phi_{\text{q}}}{L_{\text{q}}}-\frac{\dot{\Phi}_{\text{q}}}{R_{\text{q}}}+2 \frac{A_{\text{q}}^{\leftarrow }}{\sqrt{R_{\text{q}}}} ,
\\ \label{eq:classical_EOM_Q_m}
    \dot{Q}_{\text{m}} 
    &= \frac{\Phi_{\text{m}}}{L_{\text{m}}} ,
\\ \label{eq:classical_EOM_Phi_m}
    \dot{\Phi}_{\text{m}}
    & = V_{\text{m}}
    = -\frac{Q_{\text{m}}}{C_{\text{m}}}-\dot{\Phi}_{\text{q}}-R_{\text{m}} \frac{\Phi_{\text{m}}}{L_{\text{m}}}+2 \sqrt{R_{\text{m}}} A_{\text{m}}^{\rightarrow }.
\end{align}

\subsubsection{Quantization and Langevin equations}
To quantize the circuit, we define the complex amplitude
\begin{align} 
    \alpha_{\text{q}}
    &=\frac{1}{\sqrt{2 \hbar Z_{\text{q}}}}\left(\Phi_{\text{q}}+i Z_{\text{q}} Q_{\text{q}}\right), 
    \quad \alpha_{\text{q}}^*
    =\frac{1}{\sqrt{2 \hbar Z_{\text{q}}}}\left(\Phi_{\text{q}}-i Z_{\text{q}} Q_{\text{q}}\right), 
    \\
    \alpha_{\text{m}}
    &=\frac{1}{\sqrt{2 \hbar Z_{\text{m}}}}\left(\Phi_{\text{m}}-i Z_{\text{m}} Q_{\text{m}}\right), 
    \quad
    \alpha_{\text{m}}^*
    =\frac{1}{\sqrt{2 \hbar Z_{\text{m}}}}\left(\Phi_{\text{m}}+i Z_{\text{m}} Q_{\text{m}}\right),
\end{align}
where $Z_{\text{q}} = \sqrt{L_{\text{q}}/C_{\text{q}}}$ and $Z_{\text{m}} = \sqrt{L_{\text{m}}/C_{\text{m}}}$. The input traveling waves for the qubit and the mechanical resonator are normalized by the photon and phonon energy, respectively:
\begin{equation} \label{eq:traveling_wave_annihilation_ops}
    A_{\text{q}}^{\leftarrow }
    = \sqrt{\frac{\hbar \omega_{\text{q}}}{2}} \alpha_{\text{q}}^{\leftarrow },
    \quad 
    A_{\text{m}}^{\rightarrow }
    = \sqrt{\frac{\hbar \omega_{\text{m}}}{2}} \alpha_{\text{m}}^{\rightarrow }.
\end{equation}
Classically, $\alpha_{\text{q}}$ and $\alpha_{\text{m}}$ are complex functions, while $\alpha_{\text{q}}^{\leftarrow}$ and $\alpha_{\text{m}}^{\rightarrow}$ are real.

Using Eq.~(\ref{eq:classical_EOM_Phi_q})-(\ref{eq:traveling_wave_annihilation_ops}), we find
\begin{align}
    \dot{\alpha}_{\text{q}}
    &=\frac{1}{\sqrt{2 \hbar Z_{\text{q}}}} \dot{\Phi}_{\text{q}} + i \sqrt{\frac{Z_{\text{q}}}{2 \hbar}} \dot{Q}_{\text{q}}
\\
    &= \frac{1}{\sqrt{2 \hbar Z_{\text{q}}}} \frac{Q_{\text{q}}}{C_{\text{q}}} +i \sqrt{\frac{Z_{\text{q}}}{2 \hbar}}\left(\frac{\Phi_{\text{m}}}{L_{\text{m}}}-\frac{\Phi_{\text{q}}}{L_{\text{q}}} - \frac{Q_{\text{q}}}{R_{\text{q}} C_{\text{q}}} +2 \frac{A_{\text{q}}^{\leftarrow }}{\sqrt{R_{\text{q}}}}\right)
\\ 
    &=-i \omega_{\text{q}} \alpha_{\text{q}}-\frac{1}{2 R_{\text{q}} C_{\text{q}}}\left(\alpha_{\text{q}}-\alpha_{\text{q}}^*\right) + i \frac{\sqrt{\omega_{\text{q}} \omega_{\text{m}}}}{2} \sqrt{\frac{L_{\text{q}}}{L_{\text{m}}}} \left(\alpha_{\text{m}}+\alpha_{\text{m}}^*\right) + \frac{i}{\sqrt{R_{\text{q}} C_{\text{q}}}}\alpha_{\text{q}}^{\leftarrow } ,
\end{align}
\begin{align}
    \dot{\alpha}_{\text{m}} 
    &= \frac{1}{\sqrt{2 \hbar Z_{\text{m}}}} \dot{\Phi}_{\text{m}}-i \sqrt{\frac{Z_{\text{m}}}{2 \hbar}} \dot{Q}_{\text{m}} 
\\ 
    &= \frac{1}{\sqrt{2 \hbar Z_{\text{m}}}}\left(-\frac{Q_{\text{m}}}{C_{\text{m}}}-\frac{Q_{\text{q}}}{C_{\text{q}}} - \frac{R_{\text{m}}}{L_{\text{m}}} \Phi_{\text{m}} + 2 \sqrt{R_{\text{m}}} A_{\text{m}}^{\rightarrow }\right)
    -i \sqrt{\frac{Z_{\text{m}}}{2 \hbar}} \frac{\Phi_{\text{m}}}{L_{\text{m}}}
\\
    &=-i \omega_{\text{m}} \alpha_{\text{m}}
    - \frac{R_{\text{m}}}{2L_{\text{m}}} \left(\alpha_{\text{m}}+\alpha_{\text{m}}^*\right)
    + i \frac{\sqrt{\omega_{\text{m}} \omega_{\text{q}}}}{2} \sqrt{\frac{C_{\text{m}}}{C_{\text{q}}}} \left(\alpha_{\text{q}}-\alpha_{\text{q}}^*\right)
    + \sqrt{\frac{R_{\text{m}}}{L_{\text{m}}}} \alpha_{\text{m}}^{\rightarrow }.
\end{align}
Since there is no parametric drive that can bridge the frequency difference between $\alpha_{\text{q}}$ ($\alpha_{\text{m}}$) and $\alpha_{\text{q}}^*$ ($\alpha_{\text{m}}^*$), we can apply the rotating-wave approximation (RWA) to find the classical Langevin equations
\begin{align}
    \dot{\alpha}_{\text{q}}
    &=-i \omega_{\text{q}} \alpha_{\text{q}}
    -\frac{\kappa_{\text{q}}}{2}\alpha_{\text{q}} 
    + i \frac{\sqrt{\omega_{\text{q}} \omega_{\text{m}}}}{2} \sqrt{\frac{L_{\text{q}}}{L_{\text{m}}}} \alpha_{\text{m}} 
    + i \sqrt{\Gamma_{1, \text{q}}^{(0)}} \alpha_{\text{q}}^{\leftarrow } ,
\\
    \dot{\alpha}_{\text{m}}
    &=-i \omega_{\text{m}} \alpha_{\text{m}}
    -\frac{\kappa_{\text{m}}}{2} \alpha_{\text{m}} 
    + i \frac{\sqrt{\omega_{\text{q}} \omega_{\text{m}}}}{2} \sqrt{\frac{C_{\text{m}}}{C_{\text{q}}}} \alpha_{\text{q}} 
    + \sqrt{\kappa_{\text{m}}} \alpha_{\text{m}}^{\rightarrow } ,
\end{align}
where we have defined $\Gamma_{1, \text{q}}^{(0)} = 1/R_{\text{q}} C_{\text{q}}$ and $\kappa_{\text{m}} = R_{\text{m}} / L_{\text{m}}$. The RWA also implicitly removes the negative-frequency components from $\alpha_{\text{q}}^{\leftarrow}$ and $\alpha_{\text{m}}^{\rightarrow}$, making them non-Hermitian after quantization. Canonical quantization is performed by mapping the classical complex amplitude of the linear modes to the annihilation operators:
\begin{equation}
    \alpha_{\text{q}}
    \ \rightarrow \
    \hat{q}, 
    \quad
    \alpha_{\text{m}} 
    \ \rightarrow \
    \hat{b}, 
    \quad 
    \alpha_{\text{q}}^{\leftarrow} 
    \ \rightarrow \
    \hat{q}^{\text{in}}, 
    \quad
    \alpha_{\text{m}}^{\rightarrow} 
    \ \rightarrow \ 
    \hat{b}^{\text{in}}
\end{equation}
subject to the bosonic commutation relations $[\hat{q}, \hat{q}^\dagger] = [\hat{b}, \hat{b}^\dagger] =  1$ and $[\hat{q}^{\text{in}}(t), \hat{q}^{\text{in}\dagger}(t')] = [\hat{b}^{\text{in}}(t), \hat{b}^{\text{in}\dagger}(t')] = \delta(t-t')$. Consequently, the classical Langevin equations turn into the Heisenberg-Langevin equations for the annihilation operators:
\begin{align}
    \dot{\hat{q}}
    &=-i \omega_{\text{q}} \hat{q}
    -\frac{\kappa_{\text{q}}}{2}\hat{q} 
    + i \frac{\sqrt{\omega_{\text{q}} \omega_{\text{m}}}}{2} \sqrt{\frac{L_{\text{q}}}{L_{\text{m}}}} \hat{b}
    + i \sqrt{\Gamma_{1, \text{q}}^{(0)}} \, \hat{q}^{\text{in}} ,
\\
    \dot{\hat{b}}
    &=-i \omega_{\text{m}} \hat{b}
    -\frac{\kappa_{\text{m}}}{2} \hat{b}
    + i \frac{\sqrt{\omega_{\text{q}} \omega_{\text{m}}}}{2} \sqrt{\frac{C_{\text{m}}}{C_{\text{q}}}} \hat{q}
    + \sqrt{\kappa_{\text{m}}} \, \hat{b}^{\text{in}}.
\end{align}

Often, it is desirable to work with the Hamiltonian and the quantum state in the Schr\"{o}dinger picture. To map the equations of motion (i.e., the Heisenberg picture) to a simple coupled-mode Hamiltonian, we must have
\begin{equation}
    \sqrt{\frac{L_{\text{q}}}{L_{\text{m}}}} = \sqrt{\frac{C_{\text{m}}}{C_{\text{q}}}} 
    \quad \Longrightarrow \quad
    \omega_{\text{q}} = \omega_{\text{m}}.
\end{equation}
In other words, the circuit admits a coupled-mode description only when the qubit and mechanical resonator are sufficiently near resonance. Away from resonance, however, the interaction is weak, and any discrepancy in the coupling coefficient has a negligible effect. We therefore define the coupling coefficient as
\begin{equation} \label{eq:coupling_g_m}
    g_{\text{m}} \approx \frac{\sqrt{\omega_{\text{q}} \omega_{\text{m}}}}{2} \sqrt{\frac{L_{\text{q}}}{L_{\text{m}}}} \approx \frac{\sqrt{\omega_{\text{q}} \omega_{\text{m}}}}{2} \sqrt{\frac{C_{\text{m}}}{C_{\text{q}}}}
\end{equation}
so that the equations of motion correspond to the Hamiltonian
\begin{equation}
    \hat{H}_{\text{eff}} / \hbar 
    = \omega_{\text{q}} 
            \hat{q}^{\dagger}
            \hat{q}
        + \omega_{\text{m}}     
            \hat{b}^{\dagger} 
            \hat{b} 
        - g_{\text{m}} 
            \left(
                \hat{q}^{\dagger} \hat{b} 
                + \text{H.c.}
            \right) .
\end{equation}
For a transmon qubit (with anharmonicity $\alpha$) coupled to multiple mechanical modes, 
\begin{equation}
    \hat{H}_{\text{eff}} / \hbar 
    = \omega_{\text{q}} \hat{q}^{\dagger} \hat{q} 
    + \frac{\alpha}{2} 
        \hat{q}^{\dagger} 
        \hat{q}^{\dagger}
        \hat{q} \hat{q} 
    + \sum_{k}
        \left[ 
            \omega_{\text{m},k} 
                \hat{b}_{k}^{\dagger} 
                \hat{b}_{k} 
            - g_{\text{m},k}
                \left (
                    \hat{q}^{\dagger}
                    \hat{b}_{k}
                    + \text{H.c.}
                \right)
        \right] .
\end{equation}
The dissipation described by the Heisenberg-Langevin equations can be mapped to the master equation:
\begin{equation}
    \frac{\mathrm{d} \hat{\rho} }{\mathrm{d} t} 
    = - \frac{i}{\hbar}     
        \left[
            \hat{H}_{\text{eff}}, 
            \hat{\rho}
        \right] 
        + \Gamma_{1, \text{q}}^{(0)}    
            \mathcal{D}\left[\hat{q}\right] 
            \hat{\rho}
        + \sum_k \kappa_{\text{m},k} 
            \mathcal{D}\left[\hat{b}_{k}\right]
            \hat{\rho},
\end{equation}
where $\mathcal{D}[\hat{c}]\hat{\rho} = \hat{c}\hat{\rho} \hat{c}^{\dagger} - \{\hat{\rho},\hat{c}^{\dagger} \hat{c}\} / 2$. If we restrict the operation of the transmon qubit to the lowest two states, we can replace $\hat{q}$ by the Pauli operators and write
\begin{equation}
    \frac{\mathrm{d} \hat{\rho} }{\mathrm{d} t} 
    = - \frac{i}{\hbar}     
        \left[
            \hat{H}_{\text{qm}}, 
            \hat{\rho}
        \right] 
        + \Gamma_{1, \text{q}}^{(0)}    
            \mathcal{D}\left[\hat{\sigma}^{-}\right] 
            \hat{\rho}
        + \sum_k \kappa_{\text{m},k} 
            \mathcal{D}\left[\hat{b}_{k}\right]
            \hat{\rho},
\end{equation}
where $\hat{\sigma}^{\pm}$ are the raising and lowering operators for the two-level qubit and $\hat{H}_{\text{qm}}$ is defined by Eq.~(\ref{eq_hamiltonian}) in the main text.

Note that $K^2_{\text{t}} = C_{\text{m}} /C_{\text{idt}}$ is the electromechanical coupling coefficient for a classical bulk acoustic wave (BAW) resonator \cite{hashimoto_rfBAW}. Hence, for each coupling strength $g_{\text{m},k}$  derived from the equivalent circuit, we may introduce an effective electromechanical coupling coefficient $K^2_k$ via
\begin{equation}
    g_{\text{m},k} 
    = \frac{\sqrt{\omega_{\text{q}} \omega_{\text{m},k}}}{2} \sqrt{K^2_k}.
\end{equation}

\section{IDT Qubit $T_1$ in the Weak-Coupling Regime}

Having established the circuit-QED description of the qubit-SAW system, we now use it to evaluate the qubit energy relaxation rate. In the weak-coupling regime, the electromechanical couplings $g_{\text{m},k}$ are small enough that coherent qubit-mediated coupling between different mechanical modes can be neglected. The problem, therefore, reduces to a qubit interacting with a single mechanical mode, with annihilation operator $\hat{b}$. The extension to $N$ modes is then immediate: once the piezoelectricity-induced damping rate is obtained for one mode, the total damping is given by the sum over all modes.

The master equation for a qubit coupled to a single mechanical mode takes the form~\cite{gambetta_qubit_2006}
\begin{equation}
    \frac{\mathrm{d} \hat{\rho} }{\mathrm{d} t} 
    = - \frac{i}{\hbar}     
        \left[
            \hat{H}_{\text{qm}}, 
            \hat{\rho}
        \right] 
        + \Gamma_{1, \text{q}}^{(0)}    
            \mathcal{D}\left[\hat{\sigma}^{-}\right] 
            \hat{\rho}
        + 2\Gamma_{\phi, \text{q}}  
            \mathcal{D}\left[\hat{\sigma}^{+} \hat{\sigma}^{-}\right] 
            \hat{\rho}
        + \kappa_{\text{m}} \mathcal{D}\left[\hat{b}\right]
            \hat{\rho}.
\end{equation}
Note that we also include the pure dephasing rate $\Gamma_{\phi, \text{q}}$ of the qubit since it will limit the drop of the qubit $T_1$. The type of experiment used in this work ensures that the system has at most one quantum excitation. Under the weak-coupling assumption, we can restrict the Hilbert space to the span of three states: $\ket{1} = \ket{g,0}$, $\ket{2} = \ket{e,0}$, $\ket{3} = \ket{g,1}$, where $\{g,e\}$ denotes the qubit ground and excited states and $\{0,1\}$ indicates the lowest two Fock states of the mechanical modes. By moving to the interaction picture and writing the master equation explicitly in terms of $3 \times 3$ matrices, we find
\begin{align}
    \frac{\mathrm{d} \hat{\rho} }{\mathrm{d} t} 
    &= - i 
            \begin{pmatrix}
                0 & g_{\text{m}} \rho_{13} & g_{\text{m}} \rho_{12} - \Delta \rho_{13} 
            \\
                -g_{\text{m}} \rho_{13}^* & -g_{\text{m}} ( \rho_{23}^* - \rho_{23}) & -g_{\text{m}} (\rho_{33} - \rho_{22}) - \Delta \rho_{23}
            \\
                -g_{\text{m}} \rho_{12}^* + \Delta \rho_{13}^* & -g_{\text{m}} (\rho_{22} - \rho_{33}) + \Delta \rho_{23}^* & -g_{\text{m}} (\rho_{23} - \rho_{23}^*)
            \end{pmatrix}
\nonumber \\ \label{eq:qubit_SAWcavity_master_eqn}
    & \ \ \ \
        - \frac{\Gamma_{1, \text{q}}^{(0)}}{2}
            \begin{pmatrix}
                - 2 \rho_{22} & \rho_{12} & 0 \\
                \rho_{21} & 2 \rho_{22} & \rho_{23} \\
                0 & \rho_{32} & 0
            \end{pmatrix}
        - \Gamma_{\phi, \text{q}} 
            \begin{pmatrix}
                0 & \rho_{12} & 0 \\
                \rho_{21} & 0 & \rho_{23} \\
                0 & \rho_{32} & 0
            \end{pmatrix} 
        - \frac{\kappa_{\text{m}} }{2}
            \begin{pmatrix}
                -2 \rho_{33} & 0 & \rho_{13} \\
                0 & 0 & \rho_{23} \\
                \rho_{31} & \rho_{32} & 2 \rho_{33}
            \end{pmatrix},
\end{align}
where $\Delta = \omega_{\text{m}} - \omega_{\text{q}}$ and $\rho_{ij} = \bra{i}\hat{\rho} \ket{j}$ ($i,j \in \{1,2,3\}$).

Following a similar simplification presented in \cite{barends_Xmon_2013}, we obtain an exponential decay of the qubit population, i.e., $\rho_{22}(t) \propto \exp(- \Gamma_{1,\text{q}} t)$, where
\begin{equation} \label{eq:T1_reduction_model_derived}
    \Gamma_{1,\text{q}} 
    \approx \Gamma_{1,\text{q}}^{(0)}
        + \frac{
                2|g_{\text{m}}|^2 \Gamma_{2}
            }{
                \Gamma_{2}^2 + \Delta^2
            }
    \quad 
    \text{with} 
    \quad 
    \Gamma_{2}
    = \Gamma_{\phi,\text{q}} 
        + \frac{\Gamma_{1,\text{q}}^{(0)}}{2}
        + \frac{\kappa_{\text{m}}}{2} .
\end{equation}
Finally, we add up the contributions from each mechanical mode and obtain Eq.~(\ref{eq:T1_reduction_model}) in the main text.

\begin{figure*}[!h]
\centering
\includegraphics[width=\textwidth]{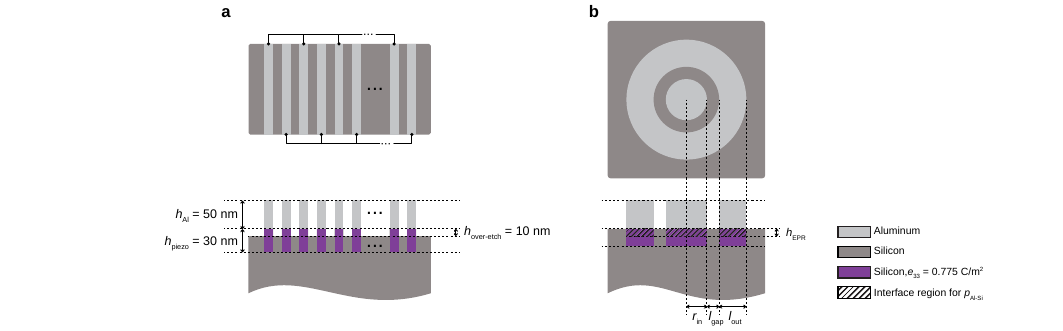}
\caption{\textbf{Schematics of the models used in the finite-element simulations}
\textbf{a}, Schematic illustration of the model used to simulate an aluminum-on-silicon interdigital capacitor/transducer. We assume that the fields in the electrical and mechanical domains have no components along the fingers, thereby reducing the problem to a two-dimensional simulation. A $h_{\text{piezo}} = 30$~nm thick silicon layer beneath the aluminum electrodes is assigned a nonzero piezoelectric coefficient $e_{33}=0.372~ \text{C}/\text{m}^2$~\cite{zhou_observation_2025}. 
During sample fabrication, silicon is overetched by $h_{\text{over-etch}}\approx 10$~nm during the formation of IDTs. This fact is included in the model.
\textbf{b}, Same as panel a, but for modeling the axially symmetric coplanar capacitor. The total energy stored in the hatched regions is used to estimate the interface energy participation ratio $p_{\text{Al-Si}}$. 
To ensure that the piezoelectric layer remains much thinner than the acoustic wavelength at high frequency, we use $h_{\text{piezo}} = 10~\text{nm}$ and $e_{33}=0.775~\text{C}/\text{m}^2$, which reproduces the experimentally obtained electromechanical coupling factor $K^2$ at 4.5~GHz~\cite{zhou_observation_2025}.
}\label{fig:S:fem_model}
\end{figure*}

\section{Simulation and IDT Qubit $T_1$ Prediction}

\begin{figure*}[b!]
\centering
\includegraphics[width=0.6\textwidth]{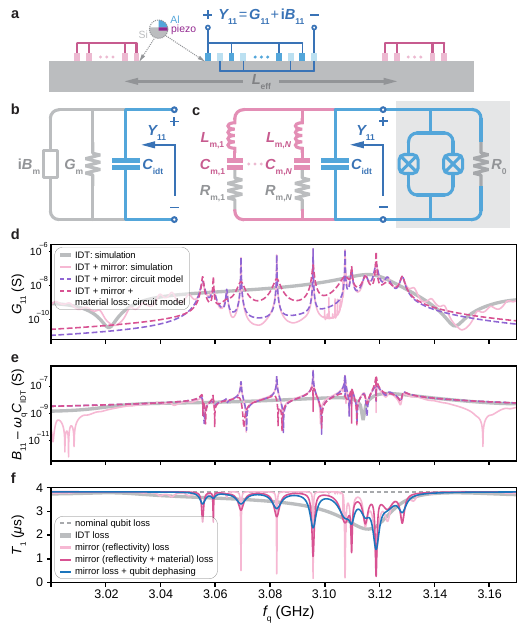}
\caption{
    \textbf{Multiphysics simulation and prediction of qubit lifetime.}
    \textbf{a}, Schematic of the simulation used to extract the effective admittance $Y_{11}$ seen by the Josephson junctions. The blue and pink rectangles represent aluminum films. A thin piezoelectric layer ($30~\text{nm}$) is inserted between the metal film and the silicon substrate to model interface-induced piezoelectricity.
    \textbf{b}, Equivalent circuit looking into the IDT port before adding the SAW reflectors (pink rectangles in \textbf{a}).
    \textbf{c}, Equivalent circuit when the SAW reflectors are included. Each SAW resonator mode is modeled as a series RLC circuit. The full qubit-SAW resonator circuit is formed by shunting the equivalent circuit with a SQUID and a resistor (for qubit nominal loss).
    \textbf{d} and \textbf{e}, Simulated $Y_{11}$ and the lumped-element fitting. 
    \textbf{f}, Predicted reductions in qubit lifetime based on black-box quantization. The nominal qubit $T_1$ (gray dotted line) is taken from the experiment (see Fig.~3f). The IDT loss (gray solid line) represents the broadband lifetime reduction due to freely propagating phonons in the absence of SAW reflectors. The mirror material loss is estimated by matching the experimentally measured mechanical quality factors of the SAW resonator modes. The blue curve additionally incorporates the experimentally characterized $T_2^*$.
}\label{fig:SI_T1_prediction}
\end{figure*}

We now turn to finite-element simulations that connect the circuit-QED description of the qubit-SAW system to the device geometry. To simulate the interface piezoelectric effect, we introduce a thin piezoelectric layer between the aluminum electrodes (Fig.~\ref{fig:S:fem_model}a) and the silicon substrate, and tune its effective piezoelectric coefficient $e_{33}$ to reproduce the IDT passband (Fig.~2f in the main text), following \cite{zhou_observation_2025}. The piezoelectric layer is assumed to share the same mechanical and electrostatic properties as silicon, except for a nonzero $e_{33}$. As shown in \cite{zhou_observation_2025}, equivalent electromechanical coupling can be reproduced using different combinations of the piezoelectric-layer thickness $h_{\text{piezo}}$ and effective coefficient $e_{33}$. These parameter choices yield similar predictions for the qubit $T_1$, indicating that the results are not sensitive to the particular parametrization of the effective piezoelectric layer. For modeling the IDT qubit, we choose $h_{\text{piezo}} = 30~\text{nm} \ll \lambda_{\text{IDT}}$, corresponding to an effective $e_{33} = 0.372 ~\text{C/m}^2$. In what follows, we validate the model by comparing the experimentally observed reduction in $T_1$ with the simulated lifetime suppression. Once validated experimentally, this workflow can be used to make quantitative estimates of the impact of interface piezoelectricity on superconducting qubits with more conventional geometries.

Establishing a connection between the classical multiphysics simulation and the prediction of $T_1$ requires a faithful mapping from the finite-element description of the device to a circuit QED model. To this end, we employ black-box quantization~\cite{Nigg_blackboxquan_2012} and extract the effective admittance $Y_{11} = G_{11} + i B_{11}$ seen by the Josephson junction in the presence of piezoelectric coupling. The two-dimensional COMSOL multiphysics simulation (schematically shown in Fig.~\ref{fig:SI_T1_prediction}a) self-consistently incorporates (i) the electrostatics of the IDT capacitor, (ii) the elastic response of the silicon substrate and aluminum electrodes, and (iii) the piezoelectric coupling at the metal-silicon interface, thereby yielding the black-box admittance at the IDT port. A perfect-matching layer (PML) ~\cite{duru_on_2025} in the mechanical domain was used at the boundary of the silicon substrate to absorb any outgoing mechanical energy flux, which corresponds to the mechanical loss of the circuit. The effective admittance can then be used to construct a lumped-element circuit representation that captures both the resonant structure and the associated dissipation (Fig.~\ref{fig:SI_T1_prediction}b,c). In the regime $\omega_{\text{q}} C_{\text{idt}} \approx \mathrm{Im}Y_{11} = B_{11}$, the qubit lifetime is given by
\begin{equation} \label{eq:idt_T1_Y_11}
    T_1(\omega_{\text{q}})
    = \frac{C_{\text{idt}}}{\mathrm{Re} Y_{11}(\omega_{\text{q}})}.
\end{equation}

In the absence of SAW reflectors, $Y_{11}$ is well described by a parallel combination of the static capacitance $C_{\text{idt}}$ and a motional admittance $G_{\text{m}} + iB_{\text{m}}$ \cite{datta_surface_1986} (Fig.~\ref{fig:SI_T1_prediction}b). The real part, $G_{\text{m}}$, corresponds to irreversible phonon radiation into the substrate and thus gives rise to broadband acoustic loss. Experimentally, this radiative decay may be difficult to distinguish from the intrinsic background loss, which we attribute to weakly coupled two-level systems and model as a resistor $R_0$ in Fig.~\ref{fig:SI_T1_prediction}c. By contrast, the introduction of SAW reflectors suppresses propagating phonon loss and places the device in a Purcell-enhanced regime that is well captured by a multimode BVD model (Fig.~\ref{fig:SI_T1_prediction}c).

Figure~\ref{fig:SI_T1_prediction}d,e shows the real ($G_{11}$) and imaginary ($B_{11}$) components of the effective admittance $Y_{11}$ obtained from the COMSOL simulation. Because $B_{11}$ is dominated by the IDT capacitance $C_{\text{idt}}$, we first extract $C_{\text{idt}}$ using a linear fit and subsequently plot only $B_{11}-\omega C_{\text{idt}}$ in order to more clearly resolve the mechanical resonances. The thick gray curves in Fig.~\ref{fig:SI_T1_prediction}d,e denote the admittance seen by the junction in the absence of SAW reflectors. In particular, the magnitude of $G_{11}$ provides a baseline for the broadband IDT emission. Upon introducing the SAW reflectors into the simulation, resonant features emerge in the extracted $Y_{11}$, shown as the solid light-pink curves in Fig.~\ref{fig:SI_T1_prediction}d,e. Because the number of periods in each reflector well exceeds that in the qubit IDT, the reflector bandwidth is narrower than the IDT passband. Consequently, away from the reflector bandwidth, $G_{11}$ follows the gray background curve.

We next fit the simulated $Y_{11}$ using the lumped-element model shown in Fig.~\ref{fig:SI_T1_prediction}c. The purple dashed lines in Fig.~\ref{fig:SI_T1_prediction}d,e reproduce the resonances (12 in total) near the center of the reflector passband and are in good agreement with the simulated $Y_{11}$. However, the COMSOL simulation does not incorporate material surface loss and therefore yields unrealistically high quality factors for the extracted mechanical modes. Since an RLC representation ($R_{\text{m},k}$, $L_{\text{m},k}$, $C_{\text{m},k}$) has already been obtained for each mechanical mode, we can adjust $R_{\text{m},k}$ to match the range of decay rates $\kappa_{\text{m},k}$ observed experimentally. This procedure yields the dark-pink dashed curves in Fig.~\ref{fig:SI_T1_prediction}d,e.

Finally, using Eq.~(\ref{eq:idt_T1_Y_11}) together with the simulated $Y_{11}$ shown in Fig.~\ref{fig:SI_T1_prediction}d,e, we estimate the reduction in qubit $T_1$. The resulting predictions under different assumptions are shown in Fig.~\ref{fig:SI_T1_prediction}f. The nominal qubit lifetime $T_1^{(0)}$ (gray dotted line in the background) is taken from experiment (see Fig.~3f in the main text). The thick gray curve represents the broadband lifetime reduction due to freely propagating phonons in the absence of SAW reflectors, whereas the light-pink curve gives the ideal reduction in $T_1$ when SAW reflectors are introduced. Analogous to the construction of the dark-pink dashed curves in Fig.~\ref{fig:SI_T1_prediction}d,e, the dark-pink curve in Fig.~\ref{fig:SI_T1_prediction}f is obtained by incorporating the experimentally extracted $\kappa_{\text{m},k}$ into the calculation, thereby reducing the predicted suppression of $T_1$.

Mechanical loss is not the only mechanism that limits the coherent electromechanical interaction. To predict the reduction in $T_1$ more accurately, one must also account for the finite qubit coherence time $T_2^*$, as captured by Eq.~(\ref{eq:T1_reduction_model}) in the main text (see also Eq.~(\ref{eq:T1_reduction_model_derived})). The electromechanical coupling coefficients $g_{\text{m},k}$ can be computed from Eq.~(\ref{eq:coupling_g_m}), since the RLC parameters of the mechanical modes have already been extracted. The resulting coupling strengths are of order $100~\text{kHz}$, in good agreement with the experimentally extracted values (Table~\ref{table:extracted_g_m}). Substituting these values into Eq.~(\ref{eq:T1_reduction_model}), instead of Eq.~(\ref{eq:idt_T1_Y_11}), yields the blue curve in Fig.~\ref{fig:SI_T1_prediction}f, in quantitative agreement with the measured reduction in $T_1$.

\section{Theoretical analysis of superconducting qubit with regular geometry}

\subsubsection{Modeling of piezoelectric loss}

\begin{figure*}[!h]
\centering
\includegraphics[width=\textwidth]{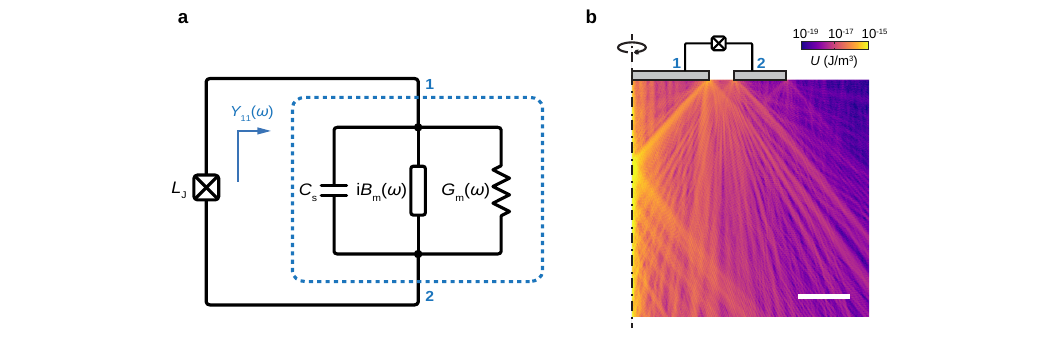}
\caption{\textbf{Simulation of interface piezoelectric loss in qubits with regular geometries.}
\textbf{a}, A more detailed circuit model of a superconducting qubit with piezoelectric loss, compared to Fig.~\ref{fig:fig4}a.
\textbf{b}, A typical mechanical energy distribution under the coplanar capacitor at $f_{\text{q}} = 4.5$~GHz. Scale bar represents 50~$\mu$m.
}\label{fig:S:qubit_loss}
\end{figure*}

The method described in the previous section is also applied to qubits with regular geometry to produce the results in Fig.~\ref{fig:fig4}.
The geometric details of the shunt capacitor of the qubit are shown in Fig.~\ref{fig:S:fem_model}b, whose single-port admittance $Y_{11}(\omega)$ is obtained from numerical simulations using the finite-element method.
A PML at the boundary of the silicon substrate is used, again, to model the irreversible phonon radiation.
A geometry with axial symmetry is used to reduce the three-dimensional problem to a two-dimensional one.
This is crucial because the electromechanical multiphysics simulation requires a mesh size significantly smaller than the acoustic wavelength to produce meaningful results.
The system's capacitance and interface energy participation ratio can be controlled by adjusting the dimensions of the aluminum electrodes ($r_{\text{in}}$, $l_{\text{gap}}$, and $l_{\text{out}}$).
We used the circuit model discussed in Ref~\cite{zhou_observation_2025}, as shown in Fig.~\ref{fig:S:qubit_loss}.
The admittance of the capacitor with the piezoelectric coupling taken into consideration (we name the system a ``lossy capacitor'' thereafter) can be expressed as
\begin{equation}
    Y_{11}(\omega) = i \omega C_{\text{s}} + G_{\text{m}} (\omega) + i B_{\text{m}} (\omega),
\end{equation}
where $C_{\text{s}}$ is the geometric shunt capacitance, $G_{\text{m}}(\omega)$ and $B_{\text{m}}(\omega)$ are the real- and imaginary-parts of the effective admittance characterizing the piezoelectric coupling and phonon radiation.
When the lossy capacitor is connected in parallel with a Josephson junction with inductance $L_{\text{J}}$, it forms an effective resonant circuit of a superconducting qubit (Fig.~\ref{fig:S:qubit_loss}a), with the resonant frequency
\begin{equation}
    f_{\text{q}} = \frac{\omega_{\text{q}}}{2\pi}\approx \frac{1}{2\pi\sqrt{L_{\text{J}} C_{\text{s}}}},
\end{equation}
and quality factor~\cite{zhou_observation_2025}
\begin{equation}
    Q_{\text{piezo}}(\omega_{\text{q}}) = \frac{{\rm Im}Y_{11}(\omega_{\text{q}})}{{\rm Re}Y_{11}(\omega_{\text{q}})}
    \approx \frac{\omega_{\text{q}} C_{\text{s}}}{G_{\text{m}}(\omega_{\text{q}})}.
\end{equation}

Fig.~\ref{fig:S:qubit_loss}c shows an example of the simulated mechanical energy density distribution inside the silicon substrate at $f_{\text{q}} = 4.5$~GHz.
The total energy is normalized to that of a single microwave photon at 4.5~GHz.
The phonon radiation-induced energy dissipation determines the real part of the effective admittance, and therefore the quality factor of the qubit.

\subsubsection{Interface energy-participation ratio and estimation of TLS-induced dielectric loss}
In general, for each qubit loss channel $L$, one may define a loss tangent $\tan \delta_L$ such that the inverse of the total qubit quality factor is
\begin{equation}
    \frac{1}{Q} 
    = \sum_{L \in \text{loss}} \frac{1}{Q_L}
    = \sum_{L \in \text{loss}} p_L \tan \delta_L,
\end{equation}
where $p_L$ denotes the energy participation ratio (EPR) of the region in which the loss mechanism resides \cite{wenner_CPW_EPR, wang_surface_EPR}. For example, the conventional interface dielectric-loss participation ratio $p_{\text{TLS}}$ includes contributions from the metal-air (MA), substrate-air (SA) and metal-substrate (MS) interfaces. Then, the TLS-limited quality factor can be calculated as
\begin{equation}
    Q_{\text{TLS}} 
    = \frac{1}{p_{\text{MA}} \tan \delta_{\text{MA}}
            + p_{\text{SA}} \tan \delta_{\text{SA}}
            + p_{\text{MS}} \tan \delta_{\text{MS}}}
    \approx \frac{1}{p_{\text{TLS}} \tan \delta_{\text{TLS}}},
\end{equation}
where we have made the common approximation that each interface has a similar loss tangent. 

Because the interface piezoelectricity originates at the MS interface, we analyze it in the main text using the aluminum-silicon interface EPR
\begin{equation}
    p_{\text{Al-Si}} 
    = \frac{E_{\text{int}}}{E_{\text{tot}}},
\end{equation}
where $E_{\text{tot}}$ is the electrostatic energy of the entire system, while $E_{\text{int}}$ is the electrostatic energy distributed within a $h_{\text{EPR}} = 3$~nm-thick silicon layer beneath the aluminum electrodes (Fig.~\ref{fig:S:fem_model}).
$h_{\text{EPR}} = 3$~nm is generally adopted in literature for estimation of the TLS dielectric loss~\cite{wang_surface_EPR}. Consequently, we compute an effective interface piezoelectric loss tangent in the main text via the expression
\begin{equation}
     \tan \delta_{\text{piezo}} 
     = \frac{1}{Q_{\text{piezo}} p_{\text{Al-Si}}}.
\end{equation}
On the other hand, for standard transmon geometries, the MS and SA interface contributions typically dominate the dielectric loss and are of comparable magnitude \cite{wenner_CPW_EPR}. We can therefore approximately relate the participation ratio for interface piezoelectricity to that of TLS as $p_{\text{Al-Si}} \approx p_{\text{TLS}}/2$.

To compare the frequency scaling of the interface piezoelectric and TLS losses, we use an empirical model of the interface TLS loss tangent~\cite{nguyen_high_2019} to capture its weak frequency dependence
\begin{equation}
    \tan\delta_{\text{TLS}}(f_{\text{q}}) = \tan\delta_{\text{TLS},0} \times \left(\frac{f_{\text{q}}}{f_{\text{q}0}}\right)^{0.15},
\end{equation}
where $f_{\text{q}0}=6$~GHz, and $\tan\delta_{\text{TLS},0} = 10^{-3}$.
We estimate the interface loss tangent to be $10^{-3}$ at 6~GHz, which is the value typically used in literature for state-of-the-art aluminum-on-silicon superconducting qubits.

We clarify that the frequency dependence of TLS loss is a complex many-body problem~\cite{guo_modeling_2026}, and the $f^{0.15}$ scaling is not expected to be accurate.
However, this weak frequency dependence is consistent with the experimental observations made in aluminum-on-silicon fluxonium qubits~\cite{nguyen_high_2019} and multiple-mode coplanar waveguide resonators~\cite{chen_broadband_2026}.
In any case, a small variation in $f_{\text{q}0}$ or in the exponent does not affect the conclusion we reached in the main text.

\end{document}